\documentclass[preprint,aps,pra,superscriptaddress,showpacs,floatfix]{revtex4}
\usepackage{graphicx}
\usepackage{wrapfig}
\usepackage{dcolumn}
\usepackage{bm}
\usepackage{amsfonts}
\usepackage{amsmath}
\usepackage{cancel}
\usepackage{multirow}
\begin{document}

\title{ Comparative study of three-nucleon force models in $A=3,4$ systems }

\author{A. Kievsky}
\author{M. Viviani}

\affiliation{
Istituto Nazionale di Fisica Nucleare, Largo Pontecorvo 3, 56127 Pisa, Italy}
\author{L. Girlanda}
\author{L.E. Marcucci}
\affiliation{ Dipartimento di Fisica, Universit\`a di Pisa,
Largo Pontecorvo 3, 56127 Pisa, Italy}

\newcommand {\nb}{\nonumber\\}
\newcommand {\he}{$^4${He} }
\newcommand {\het}{$^3${He} }
\newcommand {\trimer}{$^4${He}$_3$ }
\newcommand {\trimerecc}{$^4${He}$_3^*$ }
\newcommand {\dimer}{$^4${He}$_2$ }
\newcommand {\tetramer}{$^4${He}$_4$ }
\newcommand {\kt}{k_BT}
\newcommand {\be}{\begin{equation}}
\newcommand {\ee}{\end{equation}}
\newcommand {\bea}{\begin{eqnarray}}
\newcommand {\ea}{\end{eqnarray*}}
\newcommand {\ba}{\begin{eqnarray*}}
\newcommand {\eea}{\end{eqnarray}}
\newcommand {\ham} {{\mathcal H}}
\newcommand {\lscat} {{\mathcal L}}
\newcommand {\bra}{\langle}
\newcommand {\ket}{\rangle} 
\newcommand {\rv}{{\bf r}}
\newcommand {\tv}{{\bf t}}
\newcommand {\sv}{{\bf s}}
\newcommand {\yv}{{\bf y}}
\newcommand {\xv}{{\bf x}}
\newcommand {\xiv}{{\bf \xi}}
\newcommand {\etav}{{\bf \eta}}
\newcommand {\ab} {{\it{ab initio}}}
\newcommand {\au} {a.u.}
\newcommand {\refeq}[1] {(\ref{#1})}
\def\e{\phantom{(0)}}

\begin{abstract}
Using modern nucleon-nucleon interactions in the description of the
$A=3,4$ nuclei, it is not possible to reproduce both the three- 
and four-nucleon binding energies simultaneously. This is one manifestation of 
the necessity of including a three-nucleon force in the nuclear Hamiltonian. 
In this paper we will perform a comparative study of
some, widely used, three-nucleon force models. We will analyze their 
capability to describe the aforementioned binding energies as well as
the $n-d$ doublet scattering length. A correct description of these quantities
can be considered a stringent requirement for a nuclear
Hamiltonian containing two- and three-nucleon interaction terms.
As we will show, this requirement is not fulfilled by several of the 
models available in the literature. To satisfy it,
we propose modifications in the parametrization of the three-nucleon
forces and we study their effects on 
few selected $N-d$ low energy scattering observables.

\end{abstract}


\maketitle

\section{Introduction}
Realistic nucleon-nucleon (NN) potentials reproduce the 
experimental NN scattering data up to energies
of $350$ MeV with a $\chi^2$ per datum close to 1. 
However, the use of these potentials in the description 
of the three- and four-nucleon bound and scattering states gives a
$\chi^2$ per datum much larger than 1 (see for example Ref.~\cite{kiev01}).
In order to improve that situation, different three-nucleon
force (TNF) models have been introduced so far. Widely 
used in the literature are the Tucson-Melbourne (TM) and the 
Urbana IX (URIX) models \cite{tm,urbana}. These models are based
on the exchange mechanism of two pions between three nucleons. 
The TM model has been revisited
within a chiral symmetry approach~\cite{friar:a},
and it has been demonstrated that the contact term present in it
should be dropped. This new TM potential, known as TM', has been
subsequently readjusted~\cite{tmp}. The final operatorial structure
coincides with that one
given in the TNF of Brazil already derived many years
ago~\cite{brazil}. TNF models based on $\pi\rho$ and $\rho\rho$ meson exchange
mechanisms have also been derived~\cite{coon} and their effects have been studied in
the triton binding energy~\cite{stadler}.
More recently, TNFs have been derived~\cite{epelbaum02}
using a chiral effective field theory at next-to-next-to-leading order. 
A local version of these interactions (hereafter referred as N2LOL) 
can be found in Ref.~\cite{N2LO}. At this particular order, the TNF has
two unknown constants that have to be determined. More in general, 
all the models contain a certain number of parameters that fix the 
strength of the different terms that compose the interaction. It is a common
practice to determine these parameters from the three- and four-nucleon binding 
energies. In the chiral effective field theory there is a consistent derivation
of the two- and three-nucleon interactions and some of the low energy constants
entering in the TNF are fixed already from the NN data.
On the other hand, the parametrization of the TM' and URIX interactions
have been determined in association with specific NN potentials.
Therefore, their parametrizations could change when used with different NN potentials
since different NN potentials predict different $A=3,4$ binding energies.

The $n-d$ doublet scattering length $^2a_{nd}$ can give valuable information.
In principle this quantity is correlated, to some extent, to the $A=3$ binding
energy through the so-called Phillips line~\cite{phillips,bedaque}. 
However the presence of TNFs of the type studied here 
breaks this correlation. Therefore $^2a_{nd}$ emerges as an independent
observable that can be used to evaluate the capability of the interaction models to
describe the low energy region. Due to the lack of excited states in the
$A=3$ system, the zero energy state is the first one above the ground state. 
In the case of $n-d$ scattering at zero energy, the $J={\frac{1}{2}}^+$ state
is orthogonal to the
triton ground state and, for this reason, the wave function presents a node in the 
relative distance between the incident nucleon and the deuteron. The position of 
the node is related to the scattering length and it is also
sensitive to the relation between the overall attraction and repulsion of the
interaction. Several of the realistic NN potentials underestimate
the triton binding energy. Adding a TNF, which in general can include
an attractive as well as a repulsive component,
with a strength fixed for example to reproduce the triton binding energy, 
the balance between the overall attraction and repulsion of the interaction
changes with respect to that one produced by the NN potential alone.
And, as we will show, this leads to different predictions of $^2a_{nd}$ 
and the $\alpha$-particle binding energy
$B$($^4$He). An analysis of the parametrization of a chiral TNF,
in order to describe the triton binding energy
$B$($^3$H), $B$($^4$He) and $^2a_{nd}$, has
been performed in Ref.~\cite{epelbaum02}.
A similar analysis has not been done for the local models 
URIX, TM' and N2LOL since only the three- or four-body binding energy has been
considered in the determination of their parametrization but not $^2a_{nd}$.

In Ref.~\cite{report} results for different combinations of NN interactions
plus TNF models are given. We report the results for the quantities of
interest in Table I. From the table, we can observe that
the models are not able to describe simultaneously the $A=3,4$
binding energies and $^2a_{nd}$. Triggered by this fact,
in this paper we make a comparative study of the aforementioned TNF models.
To this end we use the AV18~\cite{av18} as the reference NN interaction
and the three-nucleon interaction models will be added to it. 
Parametrizations of the URIX and TM'
models already exist in conjunction with the AV18 potential. Conversely the
N2LOL TNF has been constructed using the N3LO-Idaho potential from Ref.
\cite{entem}. So, in a first step, we have adapted its parametrization in order
to reproduce, in conjunction with the AV18 interaction, $B$($^3$H).
Successively, we study the sensitivity of different parametrizations 
in the description of $B$($^4$He) and $^2a_{nd}$.
Selecting those parametrizations that predict these three quantities close to their
experimental values, we study some polarization observables in $p-d$ scattering
at $E_{lab}=3$ MeV. As an interesting result, we have observed that
the predictions of the different parametrizations fall in a narrow
band that, in the case of the vector analyzing powers, has a different 
position for each model, indicating a sensitivity to the
short range structure of the TNF.

All calculations have been done using the hyperspherical
harmonics (HH) method as developed by some of the authors to describe bound and
scattering states in $A=3,4$ systems~\cite{phh,kiev97,viv05,hh4s}
in configuration space or in momentum space~\cite{viv06,marcucci09}
(for a recent review see Ref.~\cite{report}).
The paper is organized as follows. In the next section we introduce the TNF models
in configuration space defining their parametrizations. In Section III we make
a sensitivity study of the parametrization for each model looking at 
$B$($^3$H), $B$($^4$He) and $^2a_{nd}$. In Section IV we study $p-d$
polarization observables at $E_{lab}=3$ MeV for specific values of the parameters.
The conclusions are given in the last section. 

\section{Three Nucleon Force Models}

In Ref.~\cite{report} the description of bound states and zero-energy
states for $A=3,4$ has been reviewed in the context of the HH method.
In Table~\ref{tb:table1} we report results for
the triton and $^4$He binding energies as well as for the doublet
$n-d$ scattering length $^2a_{nd}$ using the
AV18 and the N3LO-Idaho NN potentials and using the following
combinations of two- and three-nucleon interactions:
AV18+URIX, AV18+TM' and N3LO-Idaho+N2LOL.
The results are compared to the experimental values also reported 
in the table. Worthy of notice is
the recent  very accurate datum for $^2a_{nd}$~\cite{doublet}.

From the table we may observe that only the results obtained 
using an interaction model that includes a TNF are close to the
corresponding experimental values. In the case of the AV18+TM', the
strength of the TM' potential has been fixed to reproduce the
$^4$He binding energy and, as can be seen from the table, the 
triton binding energy is underpredicted. Conversely,
the strength of the URIX potential
has been fixed to reproduce the triton binding energy, giving too much
binding for $^4$He. The strength of the N2LOL potential has been
fixed to reproduce simultaneously the triton and the $^4$He binding
energies. In the three cases the predictions for the doublet scattering 
length are not in agreement with the experimental value, in
particular for the AV18+URIX model.

Our intention is to study different parametrizations of the TNFs to
obtain, as close as possible, a simultaneous description of the three 
quantities under observation. To this aim 
we give a brief description of the TM' (or Brazil), 
URIX and N2LOL models.  Starting from the following general TNF
\begin{equation}
W= \sum_{i<j<k} W(i,j,k)  \;\; ,
\label{eq:wijk}
\end{equation}
a generic term can be put in the following form:
\begin{equation}
W(1,2,3)= aW_a(1,2,3)+bW_b(1,2,3)+dW_d(1,2,3)+c_DW_D(1,2,3)+c_EW_E(1,2,3) \; .
\label{eq:w123}
\end{equation}
Each term corresponds to a different mechanism and has a different operatorial 
structure.
The first three terms arise from the exchange of two pions between three nucleons.
The $a$-term comes from $\pi N$ $S$-wave scattering 
whereas the $b$-term and $d$-term, which are the most important,
come from $\pi N$ $P$-wave scattering. The specific form of these three terms
in configuration space is the following:
\begin{equation}
\begin{aligned}
& W_a(1,2,3) = W_0(\bm\tau_1\cdot\bm\tau_2)(\bm\sigma_1\cdot \bm r_{31})
           (\bm\sigma_2\cdot \bm r_{23}) y(r_{31})y(r_{23}) \\
& W_b(1,2,3)= W_0 (\bm\tau_1\cdot\bm\tau_2) [(\bm\sigma_1\cdot\bm\sigma_2)
  y(r_{31})y(r_{23})  \\
 &\hspace{2cm} + (\bm\sigma_1\cdot \bm r_{31})
    (\bm\sigma_2\cdot \bm r_{23})(\bm r_{31}\cdot \bm r_{23})
  t(r_{31})t(r_{23}) \\
 & \hspace{2cm} + (\bm\sigma_1\cdot \bm r_{31})(\bm\sigma_2\cdot \bm r_{31})
  t(r_{31})y(r_{23}) \\
 &\hspace{2cm} + (\bm\sigma_1\cdot \bm r_{23})(\bm\sigma_2\cdot \bm r_{23})
  y(r_{31})t(r_{23})] \\
& W_d(1,2,3)=W_0(\bm\tau_3\cdot\bm\tau_1\times\bm\tau_2)  
   [(\bm\sigma_3\cdot \bm\sigma_2\times\bm\sigma_1)y(r_{31})y(r_{23}) \\
  & \hspace{2cm}+ (\bm\sigma_1\cdot \bm r_{31})
    (\bm\sigma_2\cdot \bm r_{23})(\bm\sigma_3\cdot\bm r_{31}\times \bm r_{23})
  t(r_{31})t(r_{23}) \\
 & \hspace{2cm} 
  + (\bm\sigma_1\cdot \bm r_{31})(\bm\sigma_2\cdot \bm r_{31}\times\bm\sigma_3)
  t(r_{31})y(r_{23}) \\
 &\hspace{2cm} + (\bm\sigma_2\cdot \bm r_{23})(\bm\sigma_3\cdot \bm r_{23}\times
  \bm\sigma_1) y(r_{31})t(r_{23})]\;\; ,
\end{aligned}
\end{equation}
with $W_0$ an overall strength.
The $b$- and $d$-terms are present in the three models whereas the $a$-term
is present in the TM' and N2LOL and not in URIX. In the first two models,
the radial functions $y(r)$ and $t(r)$ are obtained from the following function
\begin{equation}
f_0(r)=\frac{12\pi}{m_\pi^3}\frac{1}{2\pi^2}
 \int_0^\infty dq q^2 \frac{j_0(qr)}{q^2+m_\pi^2}F_\Lambda(q) 
\label{eq:f0r}
\end{equation}
where $m_\pi$ is the pion mass and
\begin{equation}
\begin{aligned}
&y(r)=\frac{1}{r} f^\prime_0(r)   \\
&t(r)=\frac{1}{r} y^\prime(r)  \,\,\ .
\end{aligned}
\label{eq:y0r}
\end{equation}
The cutoff function $F_\Lambda$ 
in the TM' or Brazil models is taken as 
$F_\Lambda=[(\Lambda^2-m_\pi^2)/(\Lambda^2+q^2)]^2$. 
In the N2LOL model it is taken as $F_\Lambda=\exp(-q^4/\Lambda^4)$. 
The momentum cutoff $\Lambda$ is a parameter of the model
fixing the scale of the problem in momentum space.
In the N2LOL, it has been fixed to $\Lambda=500$ MeV, whereas in the TM' model the ratio
$\Lambda/m_\pi$ has been varied to describe the triton or $^4$He binding energy
at fixed values of the constants $a$,$b$ and $d$. In the literature the TM' potential
has been used many times with typical 
values around $\Lambda= 5\; m_\pi$.

In the URIX model the radial dependence of the $b$- and $d$-terms is given in terms
of the functions 
\begin{equation}
\begin{aligned}
&Y(r)={\rm e}^{-x}/x\,\xi_Y   \\
&T(r)=(1+3/x+3/x^2)Y(r)\,\xi_T
\end{aligned}
\label{eq:Y0r}
\end{equation}
with $x=m_\pi r$ and the cutoff functions are defined as 
$\xi_Y=\xi_T=(1-{\rm e}^{-cr^2})$, with $c=2.1$ fm$^{-2}$. 
This regularization has been used in the AV18 potential
as well. Since the URIX model has been constructed in
conjunction with the AV18 potential, the use of the same
regularization was a choice of consistency.
The relation between the functions $Y(r),T(r)$
and those of the previous models is:
\begin{equation}
\begin{aligned}
& Y(r)=y(r)+T(r)  \\
& T(r)=\frac{r^2}{3}t(r)\,\,\, . 
\end{aligned}
\label{eq:T0r}
\end{equation}
With the definition given in Eq.(\ref{eq:f0r}), the asymptotic behavior of
the functions $f_0(r)$, $y(r)$ and $t(r)$ is:
\begin{equation}
\begin{aligned}
&f_0(r\rightarrow\infty)\rightarrow \frac{3}{m_\pi^2}\frac{{\rm e}^{-x}}{x} \\
&y(r\rightarrow\infty)\rightarrow -\frac{3{\rm e}^{-x}}{x^2}
       \left(1+\frac{1}{x}\right)  \\
&t(r\rightarrow\infty)\rightarrow \frac{3}{r^2}\frac{{\rm e}^{-x}}{x}
       \left(1+\frac{3}{x}+\frac{3}{x^2}\right) \;\; .
\end{aligned}
\label{eq:f0rasymp}
\end{equation}
To be noticed that with the normalization chosen for $f_0$, the functions $Y$ and
$T$ defined from $y$ and $t$ and those ones defined in the URIX model 
coincide at large separation distances. Conversely, they have a 
different short range behavior. Using the URIX $Y(r),T(r)$ functions, the $a$-term
has been included in the construction of the Illinois TNF model~\cite{illinois}.

The last two terms in Eq.(\ref{eq:w123}) correspond to a two-nucleon (2N) contact term
with a pion emitted or absorbed ($D$-term) and to a three-nucleon (3N)
 contact interaction ($E$-term). Their local form, in configuration space,
derived in Ref.~\cite{N2LO}, is
\begin{equation}
\begin{aligned}
& W_D(1,2,3)= W_0^D (\bm\tau_1\cdot\bm\tau_2) \{(\bm\sigma_1\cdot\bm\sigma_2)
  [y(r_{31})Z_0(r_{23})+Z_0(r_{31})y(r_{23})]  \\
 & \hspace{2cm} + (\bm\sigma_1\cdot \bm r_{31})(\bm\sigma_2\cdot \bm r_{31})
  t(r_{31})Z_0(r_{23}) \\
 &\hspace{2cm} + (\bm\sigma_1\cdot \bm r_{23})(\bm\sigma_2\cdot \bm r_{23})
  Z_0(r_{31})t(r_{23})\} \\
& W_E(1,2,3) = W_0^E(\bm\tau_1\cdot\bm\tau_2) Z_0(r_{31})Z_0(r_{23})  \,\, .
\end{aligned}
\end{equation}
The constants $W_0^D$ and $W_0^E$ fix the strength of these terms.
In the case of the URIX model the $D$-term is absent whereas 
the $E$-term is present without the isospin operatorial
structure and it has been included as purely phenomenological, without
justifying its form from a particular exchange mechanism. Its radial dependence
has been taken as $Z_0(r)=T^2(r)$.
In the N2LOL model, the function $Z_0(r)$ is defined as
\begin{equation}
Z_0(r)=\frac{12\pi}{m_\pi^3}\frac{1}{2\pi^2}
 \int_0^\infty dq q^2 j_0(qr) F_\Lambda(q)
\label{eq:z0r}
\end{equation}
with the same cutoff function used before, $F_\Lambda(q)=\exp(-q^4/\Lambda^4)$.
In the TM' model the $D$- and $E$-terms are absent.

Each model is now identified from the values assigned to the different
constants.
Following Refs.~\cite{tmp,nogga02}, in the case of the TM' model,
the values of the constants are $a=-0.87\; m^{-1}_\pi$,
$b=-2.58\; m^{-3}_\pi$, and $d=-0.753\; m^{-3}_\pi$; the strength 
$W_0=(gm_\pi/8\pi m_N)^2\;m_\pi^4$ and
the cutoff has been fixed to $\Lambda=4.756\;m_\pi$ in order to describe
correctly, associated to AV18, $B$($^4$He).
In Table~\ref{tb:table1} the calculations have been 
done using these values with $g^2=197.7$, $m_\pi=139.6$ MeV, 
$m_N/m_\pi=6.726$ ($m_N$ is the nucleon mass)
as given in the original derivation of the TM potential.

In the URIX model the $b$- and $d$-terms are present, however with
a fixed ratio based on the Fujita-Miyazawa diagram. 
The strength of these terms are:
$bW_0=4\;A^{PW}_{2\pi}$ and $d=b/4$, with $A^{PW}_{2\pi}=-0.0293$ MeV.
The model includes a purely central repulsive term introduced to
compensate the attraction of the previous term, which
by itself would produce a large overbinding in infinite nuclear matter.
It is defined as
\begin{equation}
  W_E^{URIX}(1,2,3)=A_R T^2(r_{31})T^2(r_{23})  
\end{equation}
with $A_R=0.0048$ MeV. 

In the N2LOL potential the constants of the
$a$-, $b$-, $d$-, $D$- and $E$-terms are defined in the following way:
\begin{equation}
\begin{aligned}
&  W_0=\frac{1}{12\pi^2}\left(\frac{m_\pi}{F_\pi}\right)^4g^2_A m_\pi^2  \\
&  W^D_0=\frac{1}{12\pi^2}\left(\frac{m_\pi}{F_\pi}\right)^4
        \left(\frac{m_\pi}{\Lambda_x}\right) \frac{g_A m_\pi}{8}  \\
&  W^E_0=\frac{1}{12\pi^2}\left(\frac{m_\pi}{F_\pi}\right)^4
        \left(\frac{m_\pi}{\Lambda_x}\right) m_\pi
\end{aligned}
\label{eq:constants}
\end{equation}
with $a= c_1 m^2_\pi$, $b= c_3/2$, $d= c_4/4$,  and
$c_1=-0.00081$ MeV$^{-1}$, $c_3=-0.0032$ MeV$^{-1}$, $c_4=-0.0054$ MeV$^{-1}$
taken from Ref.~\cite{entem}. The other two constants, $c_D=1.0$ and $c_E=-0.029$,
have been determined in Ref.~\cite{N2LO} from a fit to $B$($^3$H) 
and $B$($^4$He) using the N3LO-Idaho+N2LOL potential model. 
The numerical values of the constant entering in 
$W_0$, $W_0^D$ and $W_0^E$ are $m_\pi=138$ MeV, $F_\pi=92.4$ MeV, 
$g_A=1.29$, and the chiral symmetry breaking scale $\Lambda_x=700$ MeV.

\begin{figure}[b]
\begin{center}
\vspace{1cm}
\includegraphics[scale=0.6,angle=0]{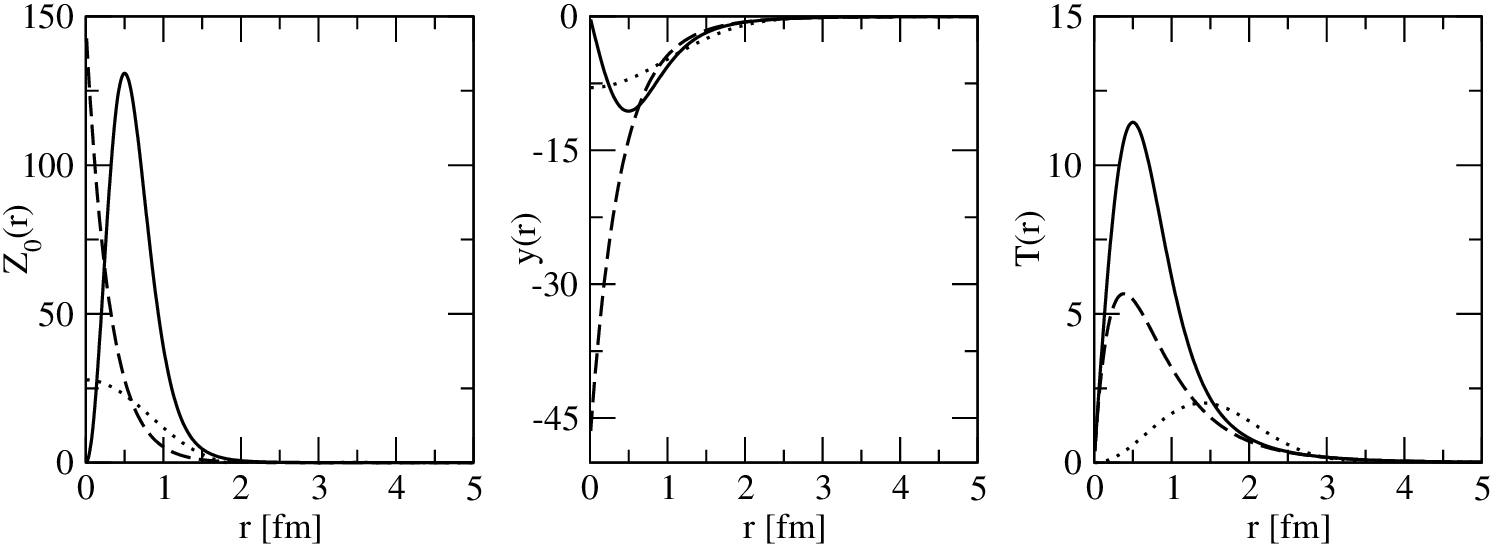}
\caption{ The $Z_0(r)$, $y(r)$ and $T(r)$ functions as functions of the
interparticle distance $r$ for the URIX (solid line), TM' (dashed line) and
N2LOL (dotted line) models.}
\label{fig:functions}
\end{center}
\end{figure}

In order to analyze the different short range structure of the TNF models,
in Fig.~\ref{fig:functions} we compare the dimensionless functions 
$Z_0(r)$, $y(r)$ and $T(r)$ for the three models under consideration.
In the TM' model using the definition
of Eq.(\ref{eq:z0r}) and using the corresponding cutoff function we can define:

\begin{equation}
 Z^{TM}_0(r)=\frac{12\pi}{m_\pi^3}\frac{1}{2\pi^2}
 \int_0^\infty dq q^2 j_0(qr) \left(\frac{\Lambda^2-m_\pi^2}{\Lambda^2+q^2}
\right)^2 
 = \frac{3}{2}\left(\frac{m_\pi}{\Lambda}\right)
   \left(\frac{\Lambda^2}{m_\pi^2}-1\right)^2 {\rm e}^{-\Lambda r}  \;\; .
\label{eq:z0rtm}
\end{equation}
This function is showed in the first panel of Fig.~\ref{fig:functions}
as a dashed line.
From the figure we can see that, in the case of the URIX model, the functions
$Z_0(r)$ and $y(r)$ go to zero as $r\rightarrow 0$. This is not the case for the
other two models and is a consequence of the regularization choice of the $Y$
and $T$ functions adopted in the URIX.

\section{Parametrization Study of the Three Nucleon Forces}

In this section we study possible variations to the parametrization of
the TNF models in order to describe the $A=3,4$ binding energies and
$^2a_{nd}$. 

\subsection{Tucson-Melbourne Force}

\begin{figure}[htb]
\begin{center}
\vspace{1.5cm}
\includegraphics[scale=0.6,angle=0]{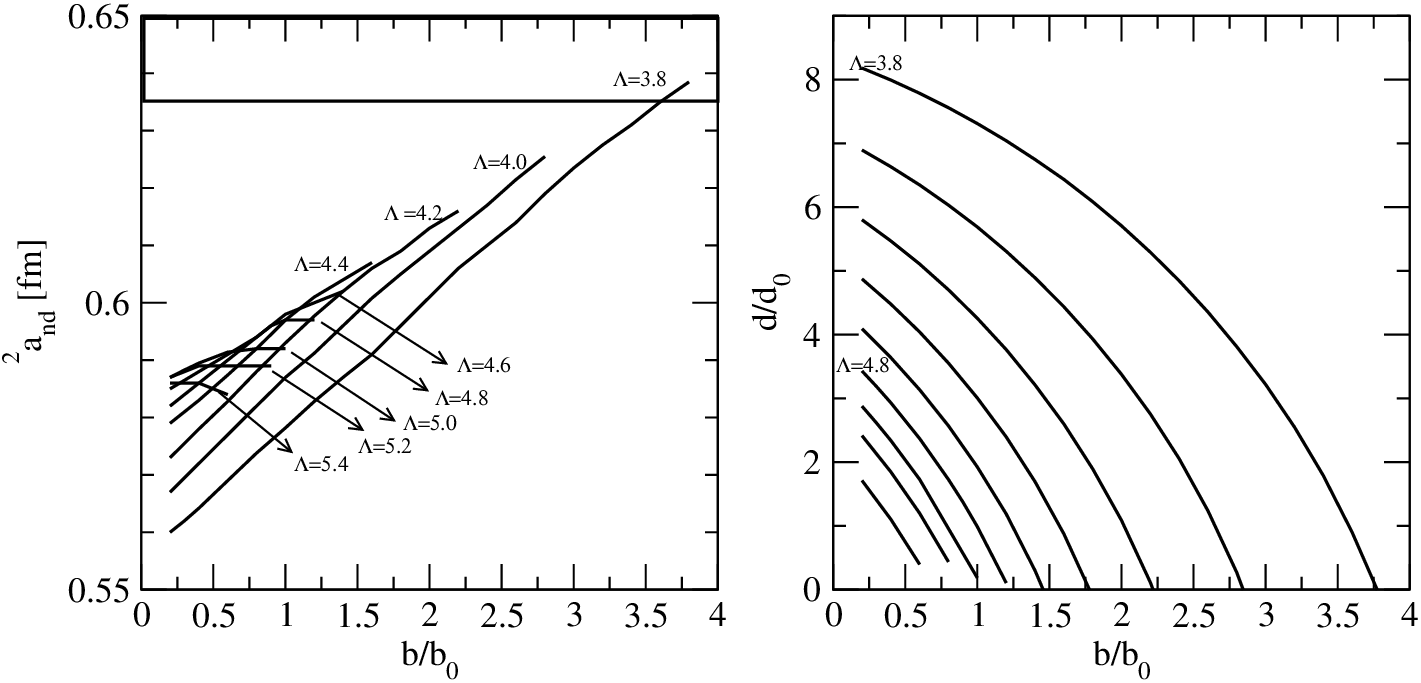}
\caption{ The doublet scattering length $^2a_{nd}$ as a function of
the parameter $b$ (in units of the original parameter $b_0=-2.58\; m_\pi^{-3}$)
of the TM' potential for different values
of the cutoff and the corresponding values of the parameter $d$
(in units of the original parameter $d_0=-0.753\; m_\pi^{-3}$),
as a function of
the parameter $b$, used to reproduce the triton binding energy.}
\label{fig:tucson}
\end{center}
\end{figure}

We first study the TM' potential and
we would like to see whether, using the AV18+TM' interaction, it is possible
to reproduce simultaneously the triton binding energy and the
doublet $n-d$ scattering length for some values of the parameters.
The $a$-term gives a very small contribution to these quantities,
therefore, in the following analysis we maintain it fixed at the value
$a=-0.87\; m^{-1}_\pi$. The analysis is shown
in Fig.~\ref{fig:tucson}. In the left panel, the doublet $n-d$ scattering length,
$^2a_{nd}$, is given as a function of the parameter $b$ (in units of
its original value $b_0=-2.58\; m^{-3}_\pi$) for
different values of the cutoff $\Lambda$ (in units of $m_\pi$). The box
in the figure includes those values of $^2a_{nd}$
compatible with the experimental results. On each point of the curves,
the value of the constant $d$ has been varied to reproduce the triton
binding energy. Its corresponding values 
(in units of its original value $d_0=-0.753\; m^{-3}_\pi$)
are given in the right panel as a function of $b$. Therefore,
each point of the curves in both panels
corresponds to a set of parameters that, in connection with the AV18 potential,
reproduces the triton binding energy.
The variations of the parameters given in Fig.~\ref{fig:tucson} do not
exhaust all the possibilities. The analysis has been done maintaining the attractive
character of the $b$- and $d$-terms and, therefore, the lines in the left
panels of the figure stop when one of the two parameters, $b$ or $d$ changes
sign. We can observe that,
with the AV18+TM' potential, there is a very small region in the parameters'
phase space available for a simultaneous description of
the triton binding energy and the doublet scattering length.
This small region corresponds to a value of $b$ around four times bigger than the
original value $b_0$ and $d$ results to be
almost zero. Moreover, the value of the cutoff $\Lambda$ around $3.8m_\pi$ is smaller
than the values usually used with the TM' potential ($\Lambda\approx5m_\pi$).

To be noticed that, for negative values of
the parameters $a$, $b$ and $d$, the TM' potential is attractive and it
does not include explicitly a repulsive term. Added to a specific NN
potential that underestimates the three-nucleon binding energy,
it supplies the extra binding by fixing appropriately its strength.
As mentioned in Sec. I, the scattering length is sensitive to
the balance between the attractive part and the repulsive part of the
complete interaction. Therefore, in the case of the TM' potential,
it seems that introducing only 
attractive terms, fixed to reproduce the triton binding energy, 
it is difficult to reproduce correctly this balance.

\begin{figure}[htb]
\begin{center}
\vspace{-0.5cm}

\includegraphics[width=14cm,height=6cm,angle=0]{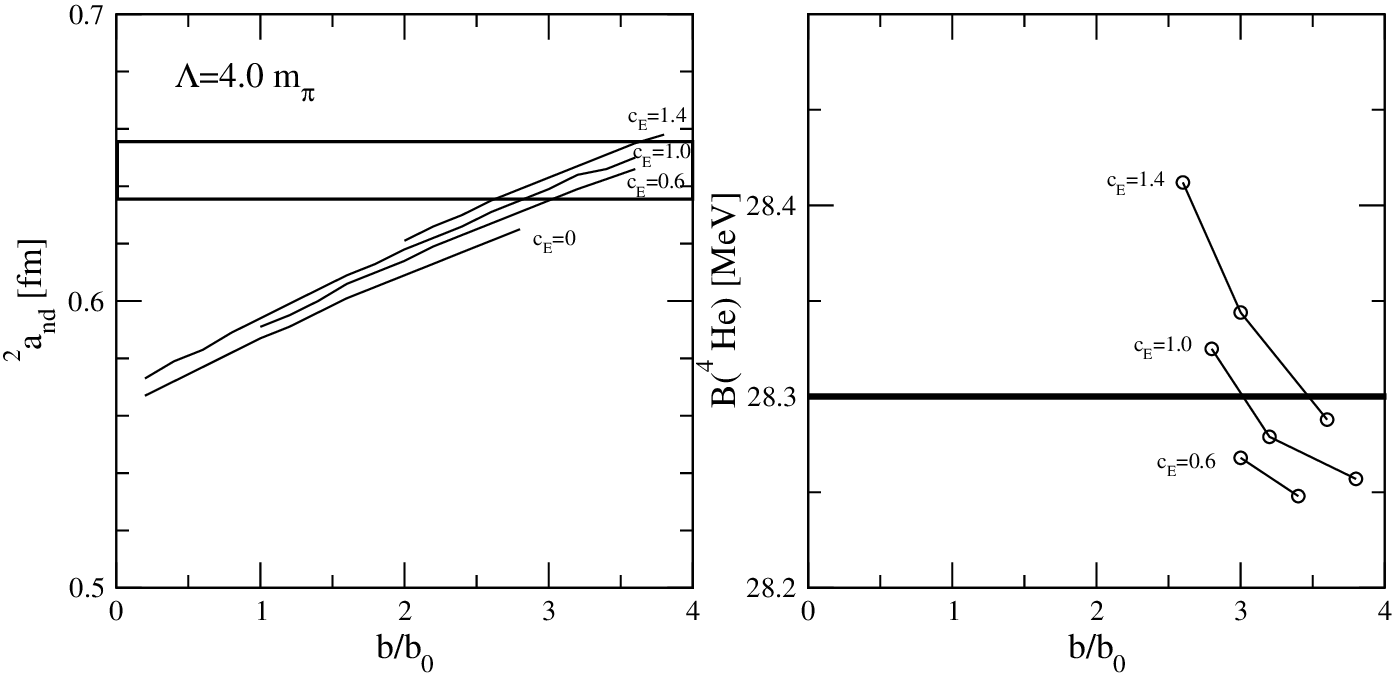}
\vspace{1.6cm}

\includegraphics[width=14cm,height=6cm,angle=0]{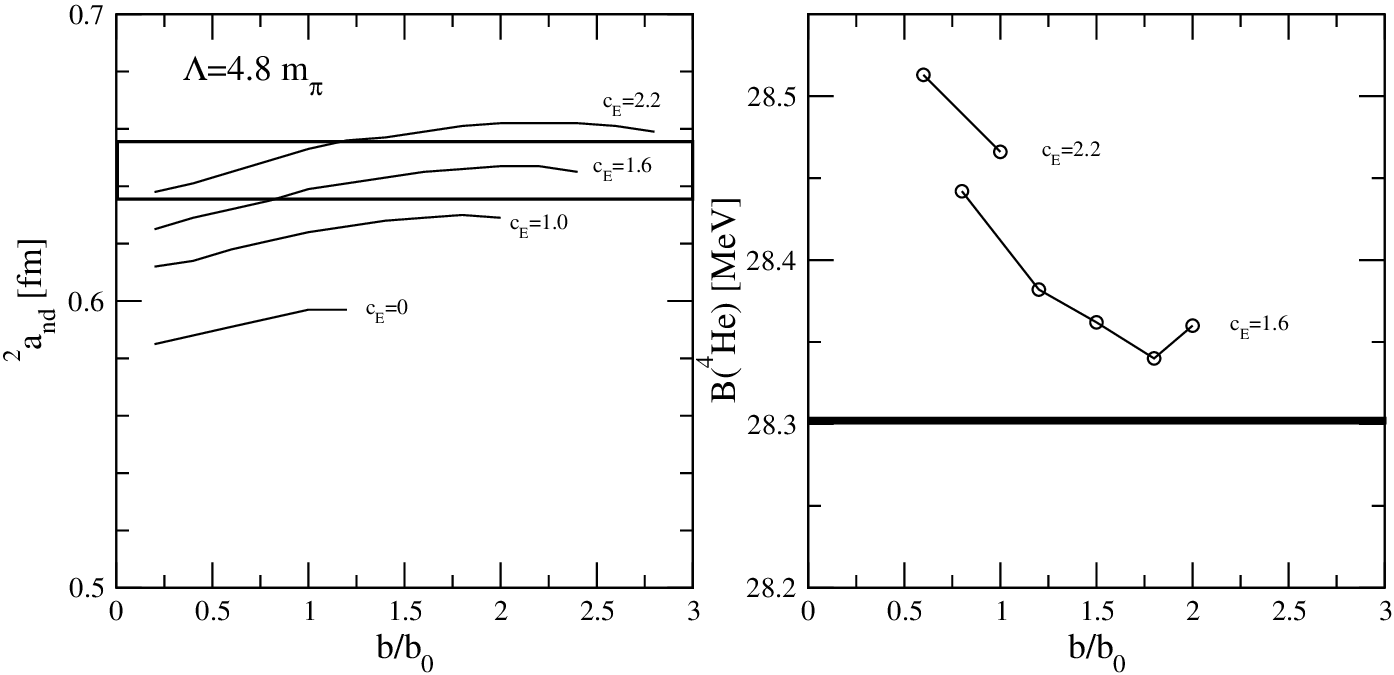}
\vspace{1.6cm}

\includegraphics[width=14cm,height=6cm,angle=0]{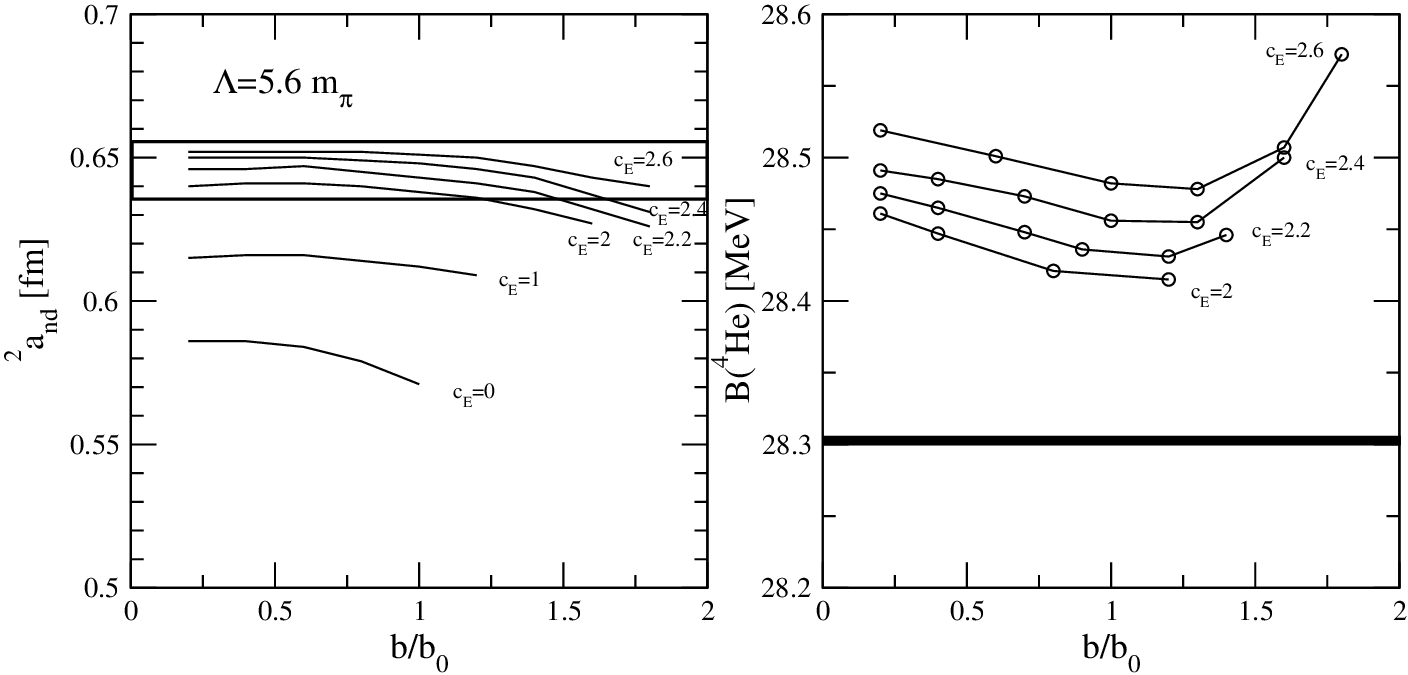}
\caption{ The doublet scattering length $a_{nd}$ as a function of
the parameter $b$ (in units of $b_0=-2.58\; m_\pi^{-3}$)
of the TM' potential including the $W_E^{TM}$-term,
for different values of the strength $c_E$ and for three selected
values of $\Lambda$. The corresponding values of $B(^4{\rm He})$,
for specific values of $a_{nd}$ inside the box, are shown in
the right panels (circles).}
\label{fig:tucson1}
\end{center}
\end{figure}

As discussed before, the TM' potential is a modification of the original
TM potential compatible with chiral symmetry. At next-to-next-to-leading order 
in the chiral effective field theory
the $D$- and $E$-terms appear (see Ref.~\cite{epelbaum02} and references therein)
as given in Eq.(\ref{eq:w123}).
Here we introduce the following additional term to the TM' potential
based on a contact term of three nucleons
\begin{equation}
W^{TM}_E(1,2,3)=W_0^E\;Z^{TM}_0(r_{31})Z^{TM}_0(r_{23})  \,\, .
\end{equation}
This term corresponds to the $E$-term in Eq.~\refeq{eq:w123}, except 
that, for the sake of simplicity,
we have omitted the $({\bm \tau}_1\cdot{\bm \tau}_2)$ operator.
Its strength $W_0^E$ is defined in Eq.~\refeq{eq:constants} and
the function $Z_0^{TM}$, defined in Eq.~\refeq{eq:z0rtm}, 
is a positive function, therefore, for positive values of
$c_E$, the new term is repulsive. We include it in the following
analysis of the TM' potential.
The results are shown in Fig.~\ref{fig:tucson1} for three
values of $\Lambda/m_\pi=4,4.8,5.6$. In the left panels
the doublet $n-d$ scattering length
is given as a function of the parameter $b$ (in units of $b_0$) for
different values of the strength $c_E$ of the $E$-term. 
The box in the panels includes those values compatible with the experimental results.
At each point of the curves, the value of the constant $d$ has been varied
to reproduce the triton binding energy. For selected values of the parameters
inside the box, the predictions for the $^4$He 
binding energy, $B(^4{\rm He})$, are shown in the right panels.

Comparing the left panels in Figs.~\ref{fig:tucson} and~\ref{fig:tucson1}, the
effect of the new term is clear. In Fig.~\ref{fig:tucson} we have observed that
using $\Lambda\ge 4\;m_\pi$, $^2a_{nd}$ cannot be well reproduced. Conversely,
in Fig.~\ref{fig:tucson1}, the inclusion of the new term allows for a description
of $^2a_{nd}$ using different values of the cutoff. The values of the parameter $b$
are closer to its original value as $\Lambda$ increases. Opposite to this,
the predictions of $B(^4{\rm He})$ improves as $\Lambda$ decreases. 
For example, considering the case $\Lambda=4\;m_\pi$, $^2a_{nd}$, $B(^3{\rm H})$
and $B(^4{\rm He})$ are well reproduced with $b=3.2b_0$, $d=6.2d_0$ and $c_E=1$.
With $\Lambda=4.8\;m_\pi$ the set of parameters that gives the best description of
the three quantities is $b=1.5b_0$, $d=4.5d_0$ and $c_E=1.6$.
And with $\Lambda=5.6\;m_\pi$ they are $b=0.8b_0$, $d=3d_0$ and $c_E=2$. 
Their different contributions to the triton binding energy are given in
Table~\ref{tb:tme}, where we report the mean
values of the kinetic energy and the NN potential energy
as well as the mean values of the attractive
part of the TNF, $V_A(3N)$, corresponding to the sum of the $a$, $b$ and 
$d$-terms, and the repulsive part, $V_R(3N)$, corresponding to the $c_E$ term.
The last two columns show $B$($^4$He) and $^2a_{nd}$. For the
sake of comparison, in the first row,
the original values of the parameters have been considered 
($b=b_0,d=d_0$ and $c_E=0$) with the value of the cutoff fixed to reproduce 
the triton binding energy ($\Lambda=4.8\;m_\pi$). As we can observe,
in this case $B$($^4$He) is overestimated and $^2a_{nd}$ is underestimated.
When the $E$-term is considered, the description of $B$($^4$He) improves and
it seems that a low value of $\Lambda$ is preferable. A further analysis
of these parametrizations is given Sec. IV studying some
polarization observables at low energy.

\subsection{Urbana IX Force}

\begin{figure}[h]
\begin{center}
\vspace{1.5cm}
\includegraphics[scale=0.6,angle=0]{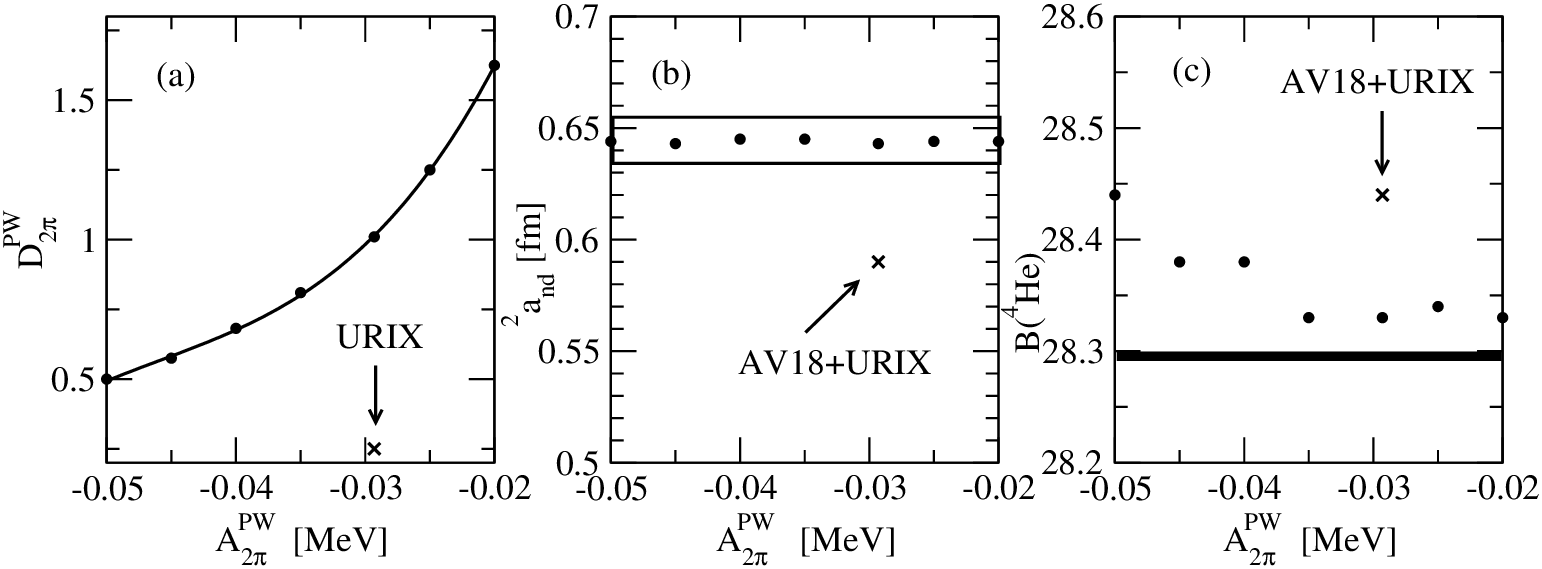}
\caption{ (a) The relative strength $D^{PW}_{2\pi}$, (b) $^2a_{nd}$
and (c) $B(^4{\rm He})$ as functions of $A^{PW}_{2\pi}$, 
for the AV18+URIX model.}
\label{fig:ur1}
\end{center}
\end{figure}

In the following we analyze the URIX potential which has two parameters, 
called $A^{PW}_{2\pi}$ and $A_R$.
In this model the strength of the $d$-term is related to the strength of
the $b$-term as $d=b/4$. The original values of the parameters have been fixed in
Ref.~\cite{urbana} in conjunction with the AV18 NN potential and, 
from Table~\ref{tb:table1}, we observe that the model correctly describes
the triton binding energy. However, it overestimates $B$($^4$He) and underestimates
$^2a_{nd}$. In order to further analyze the origin of this behavior,
we have varied the constants $A^{PW}_{2\pi}$, $A_R$ and 
the relative strength $D^{PW}_{2\pi}=d/b$ of the $b$- and $d$-terms. The regularization
parameter has been held fixed at its original value, $c=2.1$ fm$^{-2}$. 
For a given value of $A^{PW}_{2\pi}$, the values of
$A_R$ and $D^{PW}_{2\pi}$ has been chosen to reproduce $B(^3{\rm H})$ and
$^2a_{nd}$. The results are shown in Fig.~\ref{fig:ur1}. In panel (a),
$D^{PW}_{2\pi}$ is given as a function of $A^{PW}_{2\pi}$ with $A_R$
varying from $0.0176$ MeV at $A^{PW}_{2\pi}=-0.02$ MeV to
$0.0210$ MeV at $A^{PW}_{2\pi}=-0.050$ MeV. These values of $A_R$ are
more than three times bigger than the original value of $0.0048$ MeV. In panel
(b) and (c) the results for $^2a_{nd}$ and $B(^4{\rm He})$ are given
respectively. The latter has not been included in the determination
of the parameters, since $D^{PW}_{2\pi}$ and $A_R$ have been determined from
the triton binding energy and $^2a_{nd}$, and is therefore
a pure prediction. We observe a slightly overestimation
of $B(^4{\rm He})$, in particular for values of
$|A^{PW}_{2\pi}| > 0.04$ MeV, corresponding to values of $D^{PW}_{2\pi}<0.7$.

With modifications of the parameters in the URIX force, we were able to describe
reasonably well $B$($^3$H), $^2a_{nd}$ and $B$($^4$He). 
However, this has been achieved with a
substantial increase of the repulsive term. In order to gain insight on the
consequence of the new parametrizations in the quantities of interest,
in Table~\ref{tb:urbe} we report the mean
values of the kinetic energy and the NN potential energy 
as well as the mean values of the attractive
part of the TNF, $V_A(3N)$, corresponding to the sum of the $b$ and 
$d$-terms, and the repulsive part, $V_R(3N)$, corresponding to the $A_R$ term,
for selected values of the parameters (indicated as points in 
Fig.~\ref{fig:ur1}). The last two columns show 
$B$($^4$He) and $^2a_{nd}$. For the sake of comparison, in the first row,
the values obtained using the original AV18+URIX model are reported.
From the table we observe that some of the values considered for
$D^{PW}_{2\pi}$ and $A_R$ are 
quite far from the original ones. At the original value of $A^{PW}_{2\pi}$,
$-0.0293$ MeV,
the relative strength now results to be $D^{PW}_{2\pi}=1$ and  $A_R=0.0181$ MeV. As
$D^{PW}_{2\pi}$ diminishes, $A_R$ tends to increase further with the consequence
that the mean value $V_R(3N)$ is more than three times larger than the value obtained
using the original parameters (given in the first row). 
This is compensated by a lower mean value of the kinetic energy.
A further analysis of the effects of the parametrizations given in 
Table~\ref{tb:urbe} is performed in Sec. IV studying selected 
$p-d$ polarization observables.

\subsection{N2LOL Force}

\begin{figure}[h]
\begin{center}
\vspace{1.5cm}
\includegraphics[scale=0.6,angle=0]{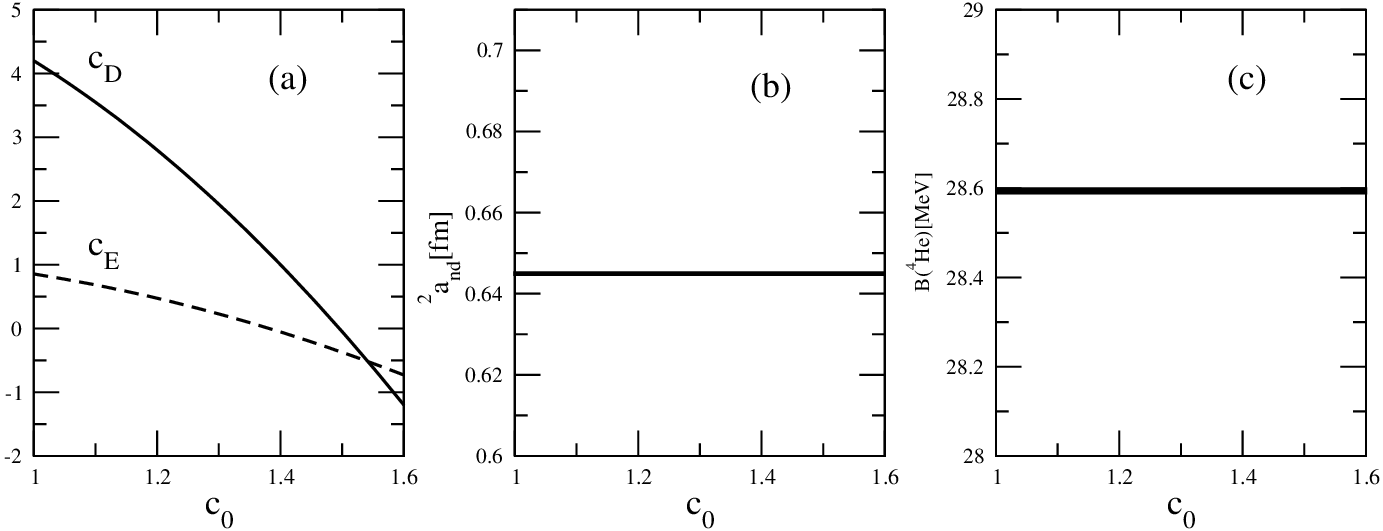}
\caption{ (a) The $c_D$ and $c_E$ parameters, (b) $^2a_{nd}$ and (c)
$B(^4{\rm He})$,  as functions of $c_0$, for the AV18+N2LOL model.}
\label{fig:n2lo}
\end{center}
\end{figure}

The parameters $c_1$, $c_3$ and $c_4$ of the N2LOL model have been taken from the
the chiral N3LO NN force of Ref.~\cite{entem}, whereas the $c_D$ and $c_E$
parameters have been determined in Ref.~\cite{N2LO}, in conjunction with
that NN force, by fitting $B$($^3$H) and
$B$($^4$He). Here we are going to use the N2LOL force in conjunction with
the AV18 NN interaction, so we have to modify its parametrization since
the amount of attraction to be gained is now different 
(see Table~\ref{tb:table1}). In the following we will call
$c^0_1$, $c^0_3$ and $c^0_4$ the values of these constants, given in Sec.II,
determined in Ref.~\cite{entem}.
Among different possibilities, in Fig.~\ref{fig:n2lo} we show a new 
parametrization of the N2LOL interaction obtained by multiplying 
$c^0_3$ and $c^0_4$ by a factor $c_0$ and maintaining $c_1=c^0_1$.
Then the parameters $c_D$ and $c_E$ have been determined from a fit to
$B$($^3$H) and $^2a_{nd}$. They are shown on panel (a)
as a function of $c_0$. Therefore, at a fixed value of $c_0$, 
with the set of parameters $c_1=c^0_1$,
$c_3=c_0c^0_3$, $c_4=c_0c^0_4$ and the 
corresponding values of $c_D$ and $c_E$ 
extracted from the figure, the AV18+N2LOL interaction
reproduces the $B$($^3$H) and $^2a_{nd}$. In panel (b) we show the stability
obtained in the description of the doublet scattering length corresponding to 
the constant value chosen for the determination of the parameters, 
$^2a_{nd}=0.644$ fm.
With the new set of parameters it is now possible to calculate $B$($^4$He).
This is shown in panel (c) and it is interesting to note that in all cases
the value $B$($^4$He)=28.60 MeV has been obtained. Modifying also the parameter
$c_1$ as $c_1=c_0c^0_1$ slightly different values of $c_D$ and $c_E$ are
obtained. Using these values to calculate $B$($^4$He), again
we obtain a constant value that is now $=28.55$ MeV.
Similar analyses using slightly different values of the cutoff $\Lambda$ around 
$500$ MeV do not change these results.

\begin{figure}[htb]
\begin{center}
\includegraphics[width=15cm,height=19cm,angle=0]{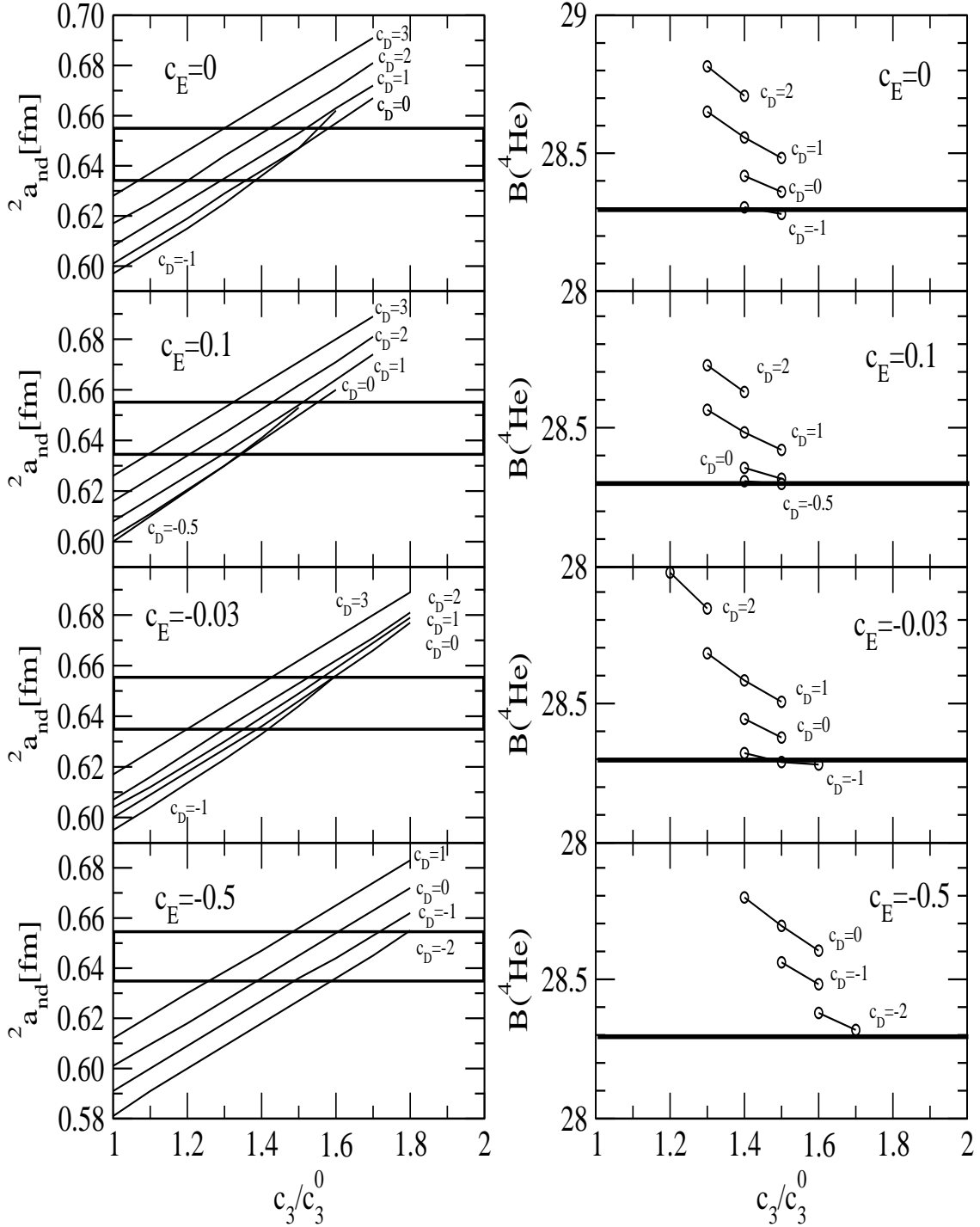}
\caption{ The doublet scattering length $^2a_{nd}$ as a function of
the parameter $c_3$ (in units of $c_3^0=-0.0032$ MeV$^{-1}$) of the N2LOL potential, 
for different values of the strength $c_D$, at four selected
values of $c_E$. The corresponding values of $B(^4{\rm He})$, for specific
values of $a_{nd}$ inside the box, are shown in the right panels (circles).}
\label{fig:n2lo1}
\end{center}
\end{figure}

In order to correctly describe $B$($^4$He), after fixing $B$($^3$H) and 
$^2a_{nd}$, we now analyze a modification to
the relative strength of the $b$- and $d$-terms which, 
in the previous analysis, was maintained at its original 
value of $c^0_4/c^0_3=1.6875$. To this end we perform a similar study
as has been done previously for the other TNF models. 
Fixing the constant $c_1$ to its original
value $c^0_1$, $c_3$, $c_4$, $c_D$ and $c_E$ have been varied. The analysis is shown 
in Fig.~\ref{fig:n2lo1} at the following four values of
$c_E=0,0.1,-0.03,-0.5$ and for the indicated values of $c_D$ chosen to
reproduce $^2a_{nd}$ (left panels). The predictions for $B$($^4$He)
are given in the right panels for those values of the parameters
that give a value of $^2a_{nd}$ inside the box.
At each point of the curves in the left panels and at the points in the right
panels, $c_4$ has been chosen to reproduce the triton binding
energy. Due to the $({\bm \tau}_i\cdot{\bm \tau}_j)$ operator in the $E$-term of the
N2LOL potential, positive values of $c_E$ makes this term attractive. Conversely,
negative values of $c_E$ makes this term repulsive. We have considered only one 
positive case, $c_E=0.1$. Increasing further $c_E$ we found it difficult to describe
correctly $B$($^4$He). For negatives values of $c_E$ we have considered two cases,
$c_E=-0.03$, which corresponds to the value given in Ref.~\cite{N2LO}, and
$c_E=-0.5$. From the figure we observe an almost linear behavior of $^2a_{nd}$.
There is a slight curvature for negative values of $c_D$ in the upper three
panels. The analysis of $B$($^4$He) selects the values of $c_D$. We have found
that the experimental value is well reproduced for the pairs 
$(c_D=-0.5,c_E=0.1)$, $(c_D=-1,c_E=0)$, $(c_D=-1,c_E=-0.03)$, 
and $(c_D=-2,c_E=-0.5)$.

In Table~\ref{tb:n2loe} we report the mean
values of the kinetic energy, the two-nucleon potential energy
as well as the mean values of the attractive part of the TNF, $V_A(3N)$, 
and its repulsive part $V_R(3N)$ for the selected values of the parameters that
correspond to the best description of the three quantities under study.
In the last two columns of the table, $B$($^4$He) and $^2a_{nd}$ are given.
The contributions to $V_A(3N)$ come
from the $a$-, $b$-, $d$ and $D$-terms, which are always
attractive in the cases considered, and from the $E$-term in the first case.
This term contributes to the repulsive part $V_R(3N)$ in the last two cases.
From the table we may observe that $c_3$ and $c_4$ results to be larger and
smaller than their original values, respectively. This is a consequence of
the simultaneous description of $B$($^3$H) and $^2a_{nd}$. Furthermore, in the
first three cases, the ratio $c_4/c_3\approx 0.46$, is
much smaller than the original ratio.

\section{Analysis of the Polarization observables}

\begin{figure}[htb]
\begin{center}
\includegraphics[width=17cm,height=18cm,angle=0]{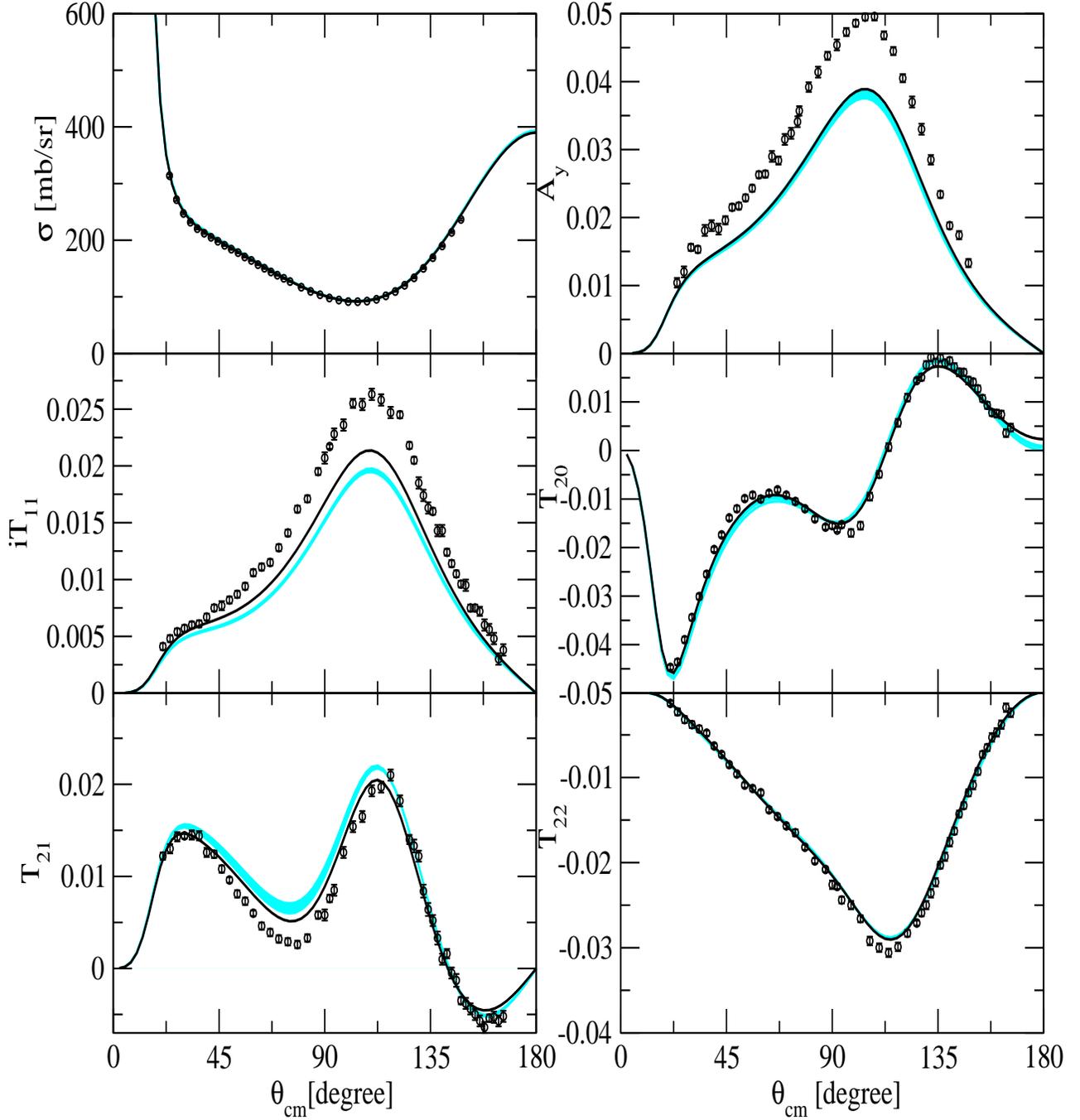}
\caption{ (Color on line)
Differential cross section and vector and tensor polarization 
observables at $E_{lab}=3$ MeV using the AV18+TM' model with the parameters 
given in the last three rows of Table~\ref{tb:tme} (cyan band). 
The predictions of the AV18+URIX
model (solid line) and the experimental points from Ref.~\protect\cite{shimizu}
are also shown.}
\label{fig:tucobs}
\end{center}
\end{figure}

\begin{figure}[htb]
\begin{center}
\includegraphics[width=17cm,height=18cm,angle=0]{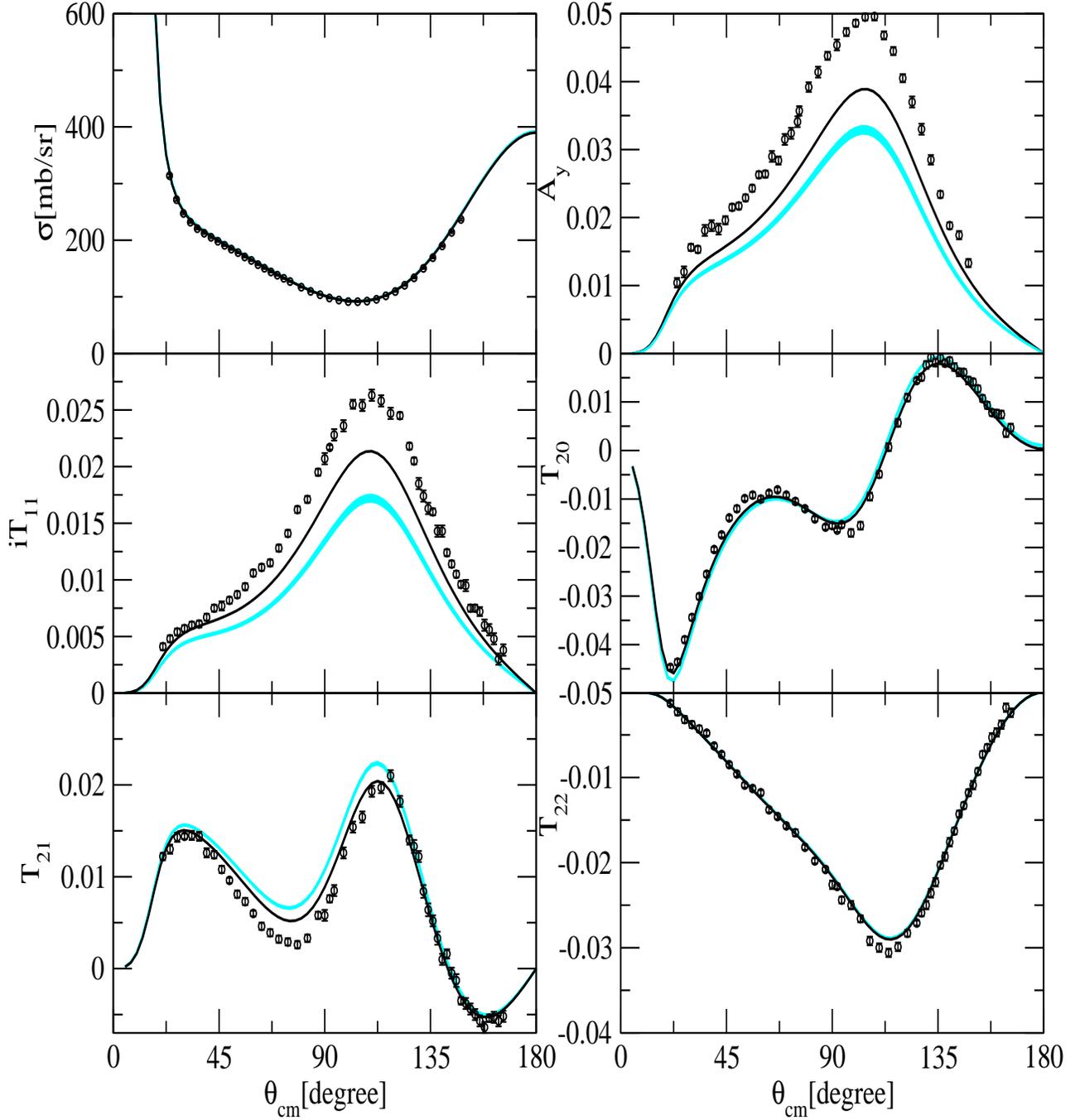}
\caption{ (Color on line)
Differential cross section and vector and tensor polarization 
observables at $E_{lab}=3$ MeV using the AV18+URIX model with the parameters 
given in Table~\ref{tb:urbe} (cyan band). The predictions of the original
AV18+URIX model, given in the first row of the table, are shown as a solid line. 
The experimental points from Ref.~\protect\cite{shimizu} are also shown.}
\label{fig:urbobs}
\end{center}
\end{figure}

\begin{figure}[htb]
\begin{center}
\includegraphics[width=17cm,height=18cm,angle=0]{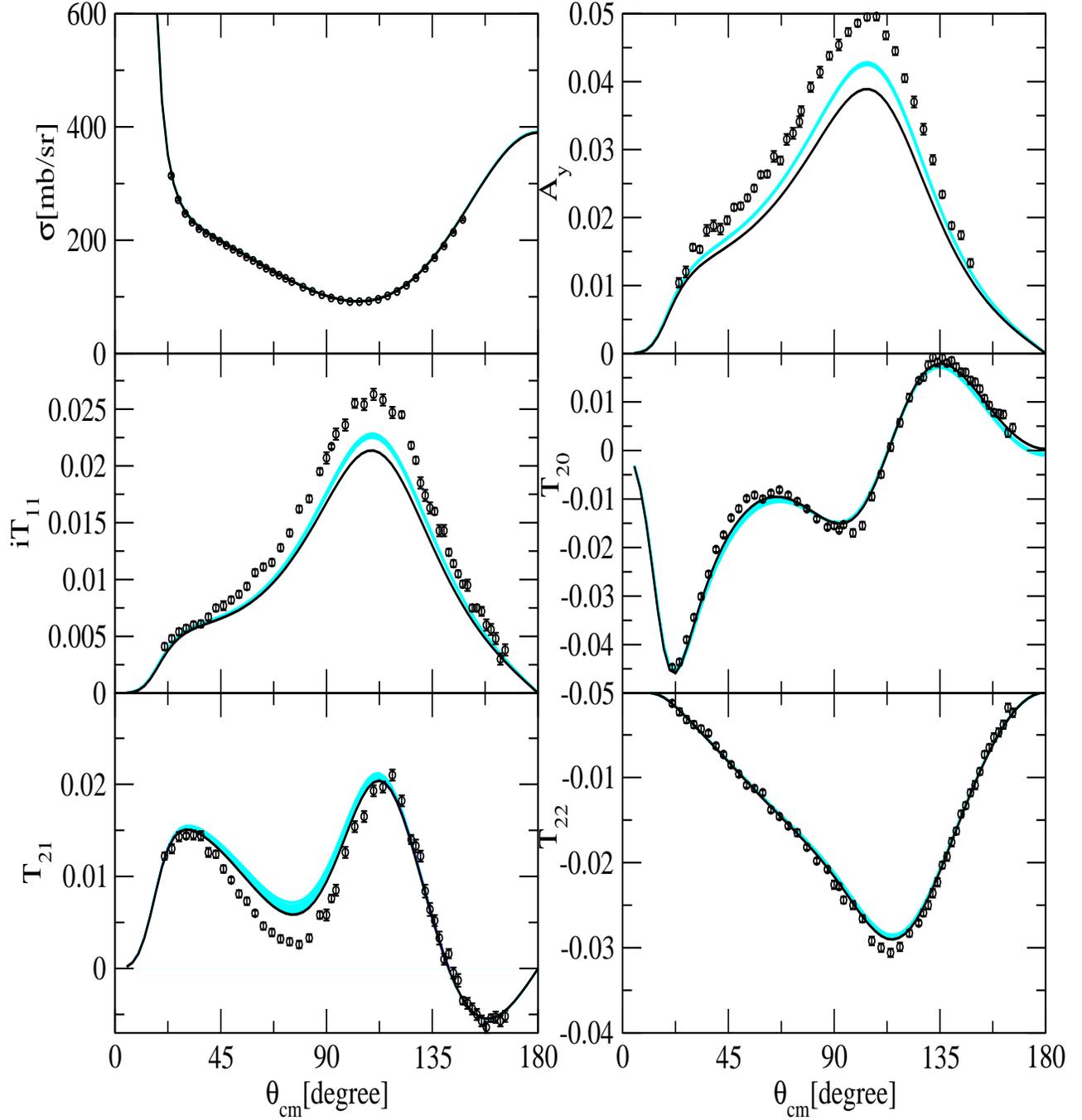}
\caption{ (Color on line)
Differential cross section and vector and tensor polarization 
observables at $E_{lab}=3$ MeV using the AV18+N2LOL model with the parameters 
given in Table~\ref{tb:n2loe} (cyan band). The predictions of the AV18+URIX
model (solid line) and the experimental points from Ref.~\protect\cite{shimizu}
are also shown.}
\label{fig:n2lobs}
\end{center}
\end{figure}

In the previous section we have studied different parametrizations of the
TM', URIX and N2LOL TNF models in conjunction with the AV18 NN potential.
The analysis has been done varying the parameters in order first to
reproduce $B$($^3$H) and then looking at their dependence 
on $^2a_{nd}$ and $B(^4{\rm He})$. To
improve the description of these quantities, some substantial modifications 
were necessary for the first two models.
In the case of the TM' interaction
we found opportune the inclusion of a repulsive term.
In the analysis of the URIX interaction, the strength of the repulsive
term resulted to be more than three times bigger than the original value and the
relative strength of the $b$- and $d$-terms, originally fixed to $1/4$, has 
been also increased. 
In the case of the N2LOL interaction, some adjustment of the parameters 
was necessary, mainly due to the fact that the AV18 interaction is
less attractive than the N3LO interaction, from which the N2LOL model
has been originally parametrized. In this section we
analyze the effects of the new parametrizations in observables that
are not correlated to the binding energies or to $^2a_{nd}$. Some polarization
observables in $p-d$ scattering have this characteristic, in particular the
vector and tensor analyzing powers. 
In Figs.~\ref{fig:tucobs},~\ref{fig:urbobs},~\ref{fig:n2lobs}
we show the differential cross section $d\sigma/d\Omega$, the vector polarization
observables $A_y$ and $iT_{11}$ and the tensor polarization observables
$T_{20}$, $T_{21}$ and $T_{22}$ at $E_{lab}=3$ MeV for the different potential
models compared to the results obtained using the original AV18+URIX interaction. 
In the figures, the cyan band collects the results obtained with the
parameters given in the three last rows of
Table~\ref{tb:tme}, from the second to the sith row of Table~\ref{tb:urbe},
and the three last rows of Table~\ref{tb:n2loe} for each model respectively,
whereas the solid line is the
prediction of the original AV18+URIX model. As we can see, for each TNF model, the
observables calculated using the different parametrizations,
fixed from a simultaneous description of
$B(^3{\rm H})$, $^2a_{nd}$ and $B(^4{\rm He})$, fall in bands which, in the case 
of the vector analyzing powers, have a different position for the three models.
Since the models essentially differ in the definitions of the 
functions $y(r)$, $T(r)$ and $Z_0(r)$, this difference can be associated to the
different short-range behavior of the TNF models.
In Fig.~\ref{fig:tucobs} we observe that the AV18+TM' model, using the
new parametrizations, does not give any
improvement in the observables compared to the AV18+URIX predictions.
Moreover $iT_{11}$ and $T_{21}$ are worse described. 
It should be observed that the AV18+TM' model, with the original parametrization,
and the AV18+URIX give similar results for the observables (a small difference
can be observed in the maximum of $A_y$ being slightly higher for the former). 
Therefore the previous
conclusions do not change if compared to the original AV18+TM' model.
In Fig.~\ref{fig:urbobs} we observe that
the new parametrizations of the AV18+URIX produce a much worse description of 
$A_y$, $iT_{11}$ and $T_{21}$. Since the vector analyzing powers are mainly
described by the $P$-wave phase-shift and mixing parameters, we can conclude that
they result to be poorly reproduced with the new parametrizations.
Conversely to what happened analyzing the previous models, 
in Fig.~\ref{fig:n2lobs}, we observe that
the N2LOL interaction produces an improvement in the description of
$A_y$ and $iT_{11}$. The well known discrepancy in these observables
is now reduced and, in particular for $A_y$, the improvement is noticeable.
In the case of the tensor
analyzing powers, a slightly worse description of $T_{21}$ between the two maxima
is now observed. In general all TNFs of the
type analyzed here have this effect in $T_{21}$ indicating that a different mechanism,
not present in the models, should be considered to improve the description of the
minimum around $75^\circ$.

\begin{figure}[htb]
\begin{center}
\includegraphics[width=17cm,height=18cm,angle=0]{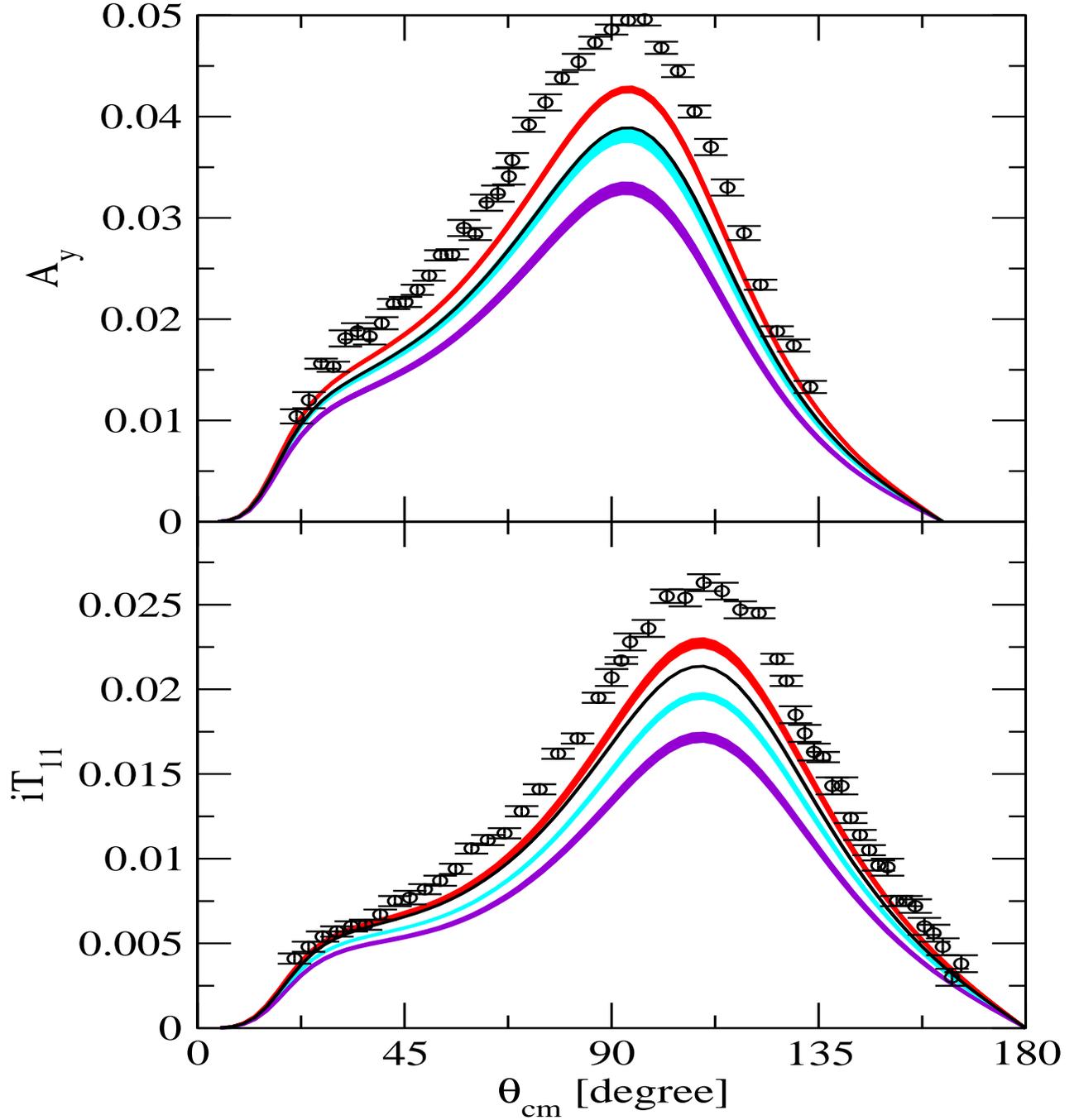}
\caption{ (Color on line) The vector analyzing powers $A_y$ and $iT_{11}$
at $E_{lab}=3$ MeV using the AV18+TM' (cyan band), AV18+URIX (violet band)
and AV18+N2LOL (red band) models 
as in Figs.~\ref{fig:tucobs},\ref{fig:urbobs},\ref{fig:n2lobs}.
The predictions of the AV18+URIX
model (solid line) and the experimental points from Ref.~\protect\cite{shimizu}
are also shown.}
\label{fig:ayfig}
\end{center}
\end{figure}

Finally we would like to comment on the fact that the vector analyzing
powers, $A_y$ and $iT_{11}$, calculated
using different TNF models fall inside a band with a different position
for each model. In Fig.~\ref{fig:ayfig} the three bands, extracted from
Figs.~\ref{fig:tucobs},\ref{fig:urbobs},\ref{fig:n2lobs}, are shown explicitly
and compared to original AV18+URIX model (solid line). We can clearly observe the
different position of the bands with the best description obtained with the new
parametrizations of the AV18+N2LO model and the worst description with those one
of the AV18+URIX model. Since all the models inside the bands describe reasonably
well $B$($^3$H), $^2a_{nd}$ and $B(^4{\rm He})$, we can conclude that the difference
is a direct consequence of their different short range structure. 
A natural question is whether, 
with opportune modifications of their radial dependence,
i.e. modifying the functions $y(r)$, $T(r)$ and $Z_0(r)$, it will be
possible to improve further the description of these observables at $E_{lab}=3$
MeV and,
eventually, obtain a $\chi^2$ per datum close to one. A preliminary study
in this direction has shown that a further improvement in the $A_y$ and
$iT_{11}$ maxima is associated to a worse description of the $T_{21}$
minimum. The particular structure of these observables is related to
a bigger splitting in the $^4P_J$ phase-shifts than the normal splitting
produced by the two-nucleon forces, as discussed in Ref.~\cite{kiev96}. 
In particular the $^4P_{1/2}$ phase-shift has to be smaller and the
mixing parameter $\epsilon_{3/2-}$ has to be bigger. It is a general
feature of the TNFs studied here that they tend to increase both,
$^4P_{1/2}$ and $\epsilon_{3/2-}$. To be more precise, in Table~\ref{tb:phases}
we show the $^4P_{J}$ phase-shifts and $\epsilon_{3/2-}$ for the AV18 and
AV18+URIX potential models, and for one selected set of the parameters of 
Tables~\ref{tb:tme},~\ref{tb:urbe},~\ref{tb:n2loe} corresponding to the new
parametrizations of the AV18+TM', AV18+URIX and AV18+N2LOL models 
(indicated in the table with an asterisk). In particular,
parametrizations of the second row of Table~\ref{tb:tme}, fourth row of 
Table~\ref{tb:urbe} and third row of Table~\ref{tb:n2loe} have been used, 
respectively.  In the last row of the
table, the results from phase-shift analysis (PSA) of Ref.~\cite{kiev96}
are given. From the table we observe that the $^4P_{1/2}$ phase-shift 
increases when the TNF models are added to the AV18 potential.
By itself this change will produce
a much worse description of $A_y$ and $iT_{11}$. However this is well compensated 
with the corresponding increase in $^4P_{5/2}$ and $\epsilon_{3/2-}$. This is not 
the case with the minimum in $T_{21}$, for which a better description would be
obtained lowering the AV18 value of $^4P_{1/2}$, as discussed in Ref.~\cite{kiev96}. 
The other parametrizations given in 
Tables~\ref{tb:tme},~\ref{tb:urbe},~\ref{tb:n2loe} produce similar changes in the
$^4P_J$ parameters.
From this observation we can conclude that the spin-isospin structure of the 
TNFs considered here is not sufficient to
describe simultaneously $B$($^3$H), $^2a_{nd}$, $B(^4{\rm He})$ and 
the vector and tensor analyzing powers at low energies.

\section{Conclusions}

Stimulated by the fact that some of the widely used TNF models 
do not reproduce simultaneously 
the triton and the $^4$He binding energies and the $n-d$ doublet scattering length, we
have analyzed possible modifications to their parametrizations. To this end
we have selected the AV18 as the reference two-nucleon force and, associated with it,
we have varied the original parameters of the TM' and URIX models so as to improve the 
description of the three quantities mentioned. Furthermore, using the recent local
form of a chiral TNF (we have called this model N2LOL), 
we have studied its parametrization associated to the AV18 interaction too. 
The analysis has
proceeded in the following way. The three models under observation, TM', AV18 and N2LOL,
have been written in configuration space as a sum of five terms, 
the $a$-, $b$-, $d$-, $C$- and $E$-terms. The
first three, corresponding to a two-pion exchange process, are attractive.
The last two, corresponding to contact terms, can be either attractive or repulsive.
Not all the models include the five terms. In the TM' model only the $a$-, $b$-
and $d$-terms are present and therefore, this model does not include explicitly
a repulsive term. 
The URIX model includes the $b$-, $d$- and $E$-terms. This last term has been
parametrized as repulsive in order to compensate the large overbinding produced by the
first two terms in infinite nuclear matter. The N2LOL model includes the five terms.  

The study has been started analyzing the AV18+TM' model. Maintaining fixed the strength of
the $a$-term at its original value, we have varied the strengths of the $b$- and $d$-terms
for several values of the cutoff parameter $\Lambda$. We have explored negative values of the
strength parameters $b$ and $d$ in order to keep the attractive character of these terms.
We have found it difficult to reproduce $^2a_{nd}$ for reasonable values of the strength 
parameters.
This fact has motivated the subsequent step of introducing a repulsive term in the model.
As a simple choice, we have introduced a purely central $E$-term and a corresponding
$Z_0(r)$ function, obtained using the monopole cutoff of the model. Including this
term we were able to describe simultaneously $B$($^3$H) and $^2a_{nd}$ for several
values of the cutoff. A further selection among these values has been done from the 
calculation of $B$($^4$He). We have observed that with $\Lambda\le 4.8\;m_\pi$ it 
was possible to describe the three quantities reasonably well. 

In the original AV18+URIX model the relative strength between the $b$- and $d$-terms 
was fixed. In the present analysis we have relaxed
this condition increasing the number of parameter of the model from two to three,
the strengths of the $b$-, $d$- and $E$-terms. Varying them,
we have found it possible to describe
the three quantities of interest for values of the parameters very different from their
original ones. In particular, the strength of the repulsive term 
resulted more than three times larger than the original value.
In the case of the AV18+N2LOL model, maintaining the strength of the $a$-term fixed
to its original value, we have varied the parameters $c_3$, $c_4$, $c_D$ and $c_E$
in combinations that reproduce $B$($^3$H). Then we have studied the dependence on
$^2a_{nd}$ and $B$($^4$He) of the different parametrizations. For fixed values
of $c_E$ we have calculated $^2a_{nd}$ for different values of $c_3$ and $c_D$.
We have found that $c_3\ge 1.4c^0_3$ in order to describe
simultaneously $B$($^3$H) and $^2a_{nd}$. The values of $c_D$ has been selected
from the analysis of $B$($^4$He). Values of $B$($^4$He) compatible with the
experimental value have been found in the four cases of $c_E$ explored.
 
After making this sensitivity study we have selected, for each model, 
some combinations of the parameters that give the better description of 
$B$($^3$H), $^2a_{nd}$ and $B$($^4$He) and we have calculated the differential
cross section and the vector and tensor analyzing powers at $E_{lab}=3$ MeV.
At this energy there are well established discrepancies between the predictions
of the theoretical models and the experimental results.
For example all potential models underestimate $A_y$ (the so-called $A_y$ puzzle)
and $iT_{11}$ and overestimate
the central minimum in $T_{21}$. Some TNF models have been constructed {\sl ad hoc}
to improve the description of these observables at low energy~\cite{kiev99}. However
the models studied here, derived from the exchange of two pions and contact terms, 
are not able to solve these discrepancies. What we have observed in the present
study is that after fixing the parameters of each model from the description
of $B$($^3$H), $^2a_{nd}$ and $B$($^4$He), the description of the vector polarization 
observables lies in a narrow band, positioned differently for each model. The best
description is given by the AV18+N2LOL model which, with respect to the original
AV18+URIX model, reduces appreciably the discrepancy in $A_y$ and $iT_{11}$.
However it gives a slightly worse description of the central minimum of $T_{21}$.
The other two models do not improve the description of the observables, compared
always to the original AV18+URIX. The modified TM' model gives similar
results though $iT_{11}$ is slightly worse whereas the results with the modified URIX
model are definitely worse than the original model. The fact that for each
model the $A_y$ and $iT_{11}$ predictions lie in a narrow band,
indicates a connection between the short-range structure of
the TNF and the polarization observables at low energies. From the analysis
we can conclude that the
smoother form of the $y(r)$, $T(r)$ and $Z_0(r)$ functions of the N2LOL potential
are preferable. To be noticed that the TM' and URIX models do not include a $D$-term.
An extended analysis of these two models including it, will allow for a more 
stringent conclusion about the short-range structure of the TNF models. 
Preliminary studies in this direction are underway.

Finally, at the end of Sec. III, we have analyzed the $^4P_J$ phase-shift parameters. 
The overall attractive character of the TNF makes larger the
$^4P_{1/2}$ and $^4P_{5/2}$ parameters, compared to the ones
obtained using a NN force alone, and has little effect on $^4P_{5/2}$. 
The mixing parameter $\epsilon_{3/2-}$ is larger too. Depending on the relative increase
of these parameters, the description of
$A_y$ and $iT_{11}$ can improve, as in the case of the N2LOL model, but not the description
of the central minimum of $T_{21}$. The spin-isospin structure of the TNF models studied here
cannot lower the $^4P_{1/2}$ phase-shift which seems to be necessary in order to improve the
description of $T_{21}$ in the minimum. A different mechanism has to be included in the
structure of the TNF, as for example it has been proposed in Ref.~\cite{kiev99}. 
Studies along this line are at present underway.

\newpage

\begin{table}[h]
\caption{The triton and $^4$He binding energies $B$ (MeV),
and  doublet scattering length $^2a_{nd}$ (fm) 
calculated using the AV18 and the N3LO-Idaho
two-nucleon potentials, and
the AV18+URIX, AV18+TM' and N3LO-Idaho+N2LOL two- and three-nucleon interactions.
The experimental values are given in the last row.}
\label{tb:table1}
\begin{tabular}{@{}llll}
\hline
Potential & $B$($^3$H) & $B$($^4$He) & $^2a_{nd}$ \cr
\hline
AV18            & 7.624    & 24.22   & 1.258 \cr
N3LO-Idaho      & 7.854    & 25.38   & 1.100 \cr
AV18+TM'        & 8.440    & 28.31   & 0.623 \cr
AV18+URIX       & 8.479    & 28.48   & 0.578 \cr
N3LO-Idaho+N2LOL & 8.474    & 28.37   & 0.675 \cr
\hline
Exp.            & 8.482    & 28.30   & 0.645$\pm$0.003$\pm$0.007 \cr
\hline
\end{tabular}
\end{table}

\begin{table}[h]
\caption{Mean values of the triton kinetic energy and the two-nucleon
potential energy $V(2N)$,
and the attractive, $V_A(3N)$, and repulsive,
$V_R(3N)$, contributions of the TNF to the triton binding energy
using the AV18+TM' potential for the specified values of the parameters
and with $a=-0.87\;m_\pi^{-1}$. In the last two columns
$B$($^4$He) and $^2a_{nd}$ are given respectively.
The experimental values are given in the last row.}
\label{tb:tme}
\begin{tabular}{@{}cccccccccc}
\hline
  $b$& $d$ & $c_E$ & $\Lambda$ & $T$ & $V(2N)$ & $V_A(3N)$ & $V_R(3N)$ &
  $B(^4$He)  & $^2a_{nd}$  \cr
  $[m^{-3}_\pi]$ & $[m^{-3}_\pi]$ &       & $[m_\pi]$ & [MeV] & [MeV] & [MeV] & [MeV]
  &  [MeV] & [fm]  \cr
\hline
-2.580 & -0.753 & 0.0 & 4.8    & 50.708 &-58.144  & -1.039 & 0.0  & 28.52 & 0.596 \cr
-8.256 & -4.690 & 1.0 & 4.0    & 50.317 &-57.366  & -2.206 & 0.781& 28.30 & 0.644 \cr
-3.870 & -3.375 & 1.6 & 4.8    & 50.699 &-57.641  & -2.748 & 1.215& 28.38 & 0.644 \cr
-2.064 & -2.279 & 2.0 & 5.6    & 50.998 &-57.940  & -2.814 & 1.291& 28.44 & 0.640 \cr
\hline
Exp.   &        &     &        &        &         &        &      & 28.30  
& 0.645$\pm$0.003$\pm$0.007 \cr
\hline
\end{tabular}
\end{table}

\begin{table}[h]
\caption{Mean values of the triton kinetic energy, the two-nucleon
potential energy $V(2N)$, and the attractive, $V_A(3N)$, and repulsive,
$V_R(3N)$, contributions of the TNF to the triton binding energy
using the AV18+URIX potential for the specified values of the parameters. 
In the last two columns 
$B$($^4$He) and $^2a_{nd}$ are given respectively.
The experimental values are given in the last row.}
\label{tb:urbe}
\begin{tabular}{@{}ccccccccc}
\hline
$A^{PW}_{2\pi}$& $D^{PW}_{2\pi}$& $A_R$ & $T$ & $V(2N)$ & $V_A(3N)$ & $V_R(3N)$ &
  $B$($^4$He) & $^2a_{nd}$ \cr
  [MeV] &  & [MeV] & [MeV] & [MeV] & [MeV] & [MeV]
  &  [MeV] & [fm]  \cr
\hline
-0.0293 & 0.25   & 0.0048  & 51.259 & -58.606 & -1.126 & 1.000 & 28.48   & 0.578 \cr
-0.0200 & 1.625  & 0.0176  & 47.472 & -57.976 & -0.923 & 2.950 & 28.33   & 0.644 \cr
-0.0250 & 1.25   & 0.0182  & 47.628 & -57.967 & -1.162 & 3.024 & 28.34   & 0.644 \cr
-0.0293 & 1.00   & 0.0181  & 47.876 & -58.000 & -1.369 & 3.015 & 28.33   & 0.643 \cr
-0.0350 & 0.8125 & 0.0191  & 47.998 & -57.975 & -1.649 & 3.147 & 28.33   & 0.645 \cr
-0.0400 & 0.6875 & 0.0198  & 48.133 & -57.964 & -1.897 & 3.249 & 28.38   & 0.645 \cr
-0.0450 & 0.5625 & 0.0198  & 48.414 & -57.995 & -2.148 & 3.248 & 28.38   & 0.643 \cr
-0.0500 & 0.50   & 0.0210  & 48.471 & -57.952 & -2.401 & 3.401 & 28.44   & 0.645 \cr
\hline
Exp.   &         &         &        &         &        &       & 28.30  
& 0.645$\pm$0.003$\pm$0.007 \cr
\hline
\end{tabular}
\end{table}

\begin{table}[h]
\caption{Mean values of the triton kinetic energy, the two-nucleon
potential energy $V(2N)$, and the attractive, $V_A(3N)$,
and repulsive, $V_R(3N)$, TNF contributions to the triton potential energy
using the AV18+N2LOL potential for the specified values of the
parameters and with $c_1=-0.00081$ MeV$^{-1}$. 
In the last two columns $B$($^4$He) and $^2a_{nd}$ are given.
The experimental values are given in the last row.}
\label{tb:n2loe}
\begin{tabular}{@{}cccccccccc}
\hline
 $c_3$  & $c_4$ & $c_D$ & $c_E$ & $T$ & $V(2N)$ & $V_A(3N)$ & $V_R(3N)$ &
  $B$($^4$He) & $^2a_{nd}$ \cr
  $[c_3^0]$ & $[c_4^0]$ &  &  & [MeV] & [MeV] & [MeV] & [MeV]
  &  [MeV] & [fm]  \cr
\hline
 1.4 & 0.3636 & -0.5& 0.1 & 49.834 & -57.278 & -1.029 & 0.0   & 28.31   & 0.641 \cr
 1.4 & 0.3786 & -1  & 0.0 & 49.950 & -57.401 & -1.022 & 0.0   & 28.30   & 0.636 \cr
 1.5 & 0.3735 & -1  &-0.03& 49.839 & -57.274 & -1.076 & 0.036 & 28.29   & 0.644 \cr
 1.7 & 0.9000 & -2  &-0.50& 50.166 & -57.181 & -2.119 & 0.657 & 28.32   & 0.645 \cr
\hline
Exp.  &       &     &     &        &         &        &       & 28.30  
& 0.645$\pm$0.003$\pm$0.007 \cr
\hline
\end{tabular}
\end{table}

\begin{table}[h]
\caption{ The $^4P_J$ phase-shifts and the $\epsilon_{3/2-}$ mixing parameter
at $E_{lab}=3$ MeV for the potential models indicated. For the sake of
comparison, the results of the
PSA from Ref.~\protect\cite{kiev96} are given in the last row. }
\label{tb:phases}
\begin{tabular}{@{}ccccc}
\hline
     & $^4P_{1/2}$ & $^4P_{3/2}$ & $^4P_{5/2}$ & $\epsilon_{3/2-}$  \cr
\hline
AV18      & 22.03 & 24.24 & 24.08 & -2.247 \cr
AV18+URIX & 22.31 & 24.30 & 24.27 & -2.314 \cr
\hline
AV18+TM'*   & 22.79 & 24.45 & 24.53 & -2.453 \cr
AV18+URIX*  & 22.75 & 24.41 & 24.35 & -2.375 \cr
AV18+N2LOL* & 22.55 & 24.25 & 24.48 & -2.394 \cr
\hline
PSA         & $21.77\pm 0.01$ & $24.30\pm 0.01$ & $24.26\pm 0.01$ & $-2.46\pm 0.01$\cr
\hline
\end{tabular}
\end{table}

\newpage

\begin{thebibliography}{9}

\bibitem{kiev01} A. Kievsky, M. Viviani, and S. Rosati, Phys. Rev. C {\bf 64},
 024002 (2001)
\bibitem{tm}S.A. Coon and W. Gl\"ockle, Phys. Rev. C {\bf 23}, 1790 (1981) 
\bibitem{urbana} B.S. Pudliner, V. R. Pandharipande, J. Carlson, and R.B. Wiringa,
Phys. Rev. Lett. {\bf 74}, 4396 (1995)
\bibitem{friar:a} J.L. Friar, D. H\"uber, and U. van Kolck,
Phys. Rev. C {\bf 59}, 53 (1999)
\bibitem{tmp} S.A. Coon  and H.K. Han,
Few-Body Syst. {\bf 30}, 131 (2001)
\bibitem{brazil} H.T. Coelho, T.K. Das, and M.R. Robilotta,
Phys. Rev. C {\bf 28}, 1812 (1983);
M.R. Robilotta and H.T. Coelho,
Nucl. Phys. A {\bf 460}, 645 (1986)
\bibitem{coon} S.A. Coon and M.T. Pe\~{n}a, Phys. Rev. C {\bf 48}, 2559 (1993)
\bibitem{stadler} A. Stadler, J. Adam, Jr., H. Henning, and
   P.U. Sauer, Phys. Rev. C {\bf 51}, 2896 (1995)
\bibitem{epelbaum02} E. Epelbaum {\sl et al.}, Phys. Rev. C {\bf 66},
 064001 (2002)
\bibitem{N2LO} P. Navratil, Few-Body Syst. {\bf 41}, 117 (2007)
\bibitem{phillips} A.C. Phillips, Nucl. Phys. A {\bf 107}, 209 (1968)
\bibitem{bedaque} P.F. Bedaque, H.-W. Hammer, and U. van Kolck, Nucl.
Phys. A {\bf 646}, 444 (1999)
\bibitem{report}A. Kievsky, S. Rosati, M. Viviani, L.E. Marcucci, and 
 L. Girlanda, J. Phys. G: Nucl. Part. Phys. {\bf 35}, 063101 (2008) 
\bibitem{av18} R.B.\ Wiringa, V.G.J.\ Stoks, and R.\ Schiavilla,
                Phys.\ Rev.\ C {\bf 51} 38 (1995)
\bibitem{entem}D.R.\ Entem and R.\ Machleidt,
                Phys.\ Rev.\ C {\bf 68} 041001(R) (2003)
\bibitem{phh}A.\ Kievsky, M.\ Viviani, and S.\ Rosati,
                Nucl.\ Phys.\ A {\bf 577}, 511 (1994)
\bibitem{kiev97} A. Kievsky, Nucl. Phys. A {\bf 624}, 125 (1997)
\bibitem{viv05}M.\ Viviani, A.\ Kievsky, and S.\ Rosati,
                Phys.\ Rev.\ C {\bf 71} 024006 (2005)
\bibitem{hh4s}R. Lazauskas {\sl et al.}, Phys.Rev. C {\bf 71} 034004 (2005)
\bibitem{viv06} M. Viviani, L.E. Marcucci, S. Rosati, A. Kievsky, and
       L. Girlanda, Few-Body Syst.{\bf 39}, 159 (2006)
\bibitem{marcucci09} L.E. Marcucci, A. Kievsky, L. Girlanda, S. Rosati and 
    M. Viviani, Phys. Rev. C {\bf 80}, 034003 (2009)
\bibitem{doublet} K. Schoen {\sl et al.}, Phys. Rev. C {\bf 67}, 044005 (2003)
\bibitem{illinois}S.C. Pieper, V.R. Pandharipande, R.B. Wiringa, and J. Carlson,
Phys. Rev. C {\bf 64}, 014001 (2001)
\bibitem{nogga02} A. Nogga, H. Kamada, W. Gl\"ockle, and B.R. Barrett,
              Phys. Rev. C {\bf 65}, 054003 (2002)
\bibitem{shimizu} S. Shimizu {\sl et al.}, Phys. Rev. C {\bf 52}, 1193 (1995)
\bibitem{kiev96} A. Kievsky, S. Rosati, W. Tornow, and M. Viviani,
        Nuc. Phys. A {\bf 607}, 402 (1996)
\bibitem{kiev99} A. Kievsky, Phys. Rev. C {\bf 60}, 034001 (1999)


\end{thebibliography}
\end{document}